\documentstyle[multicol,aps,psfig]{revtex}

\begin{document}

\title{Dynamical Systems approach to Saffman-Taylor fingering.\\
A Dynamical Solvability Scenario}

\author{
E. Paun\'e, F.X. Magdaleno and J. Casademunt
}

\address{
Departament d'Estructura i Constituents de la Mat\`eria\\
Universitat de Barcelona,
Av. Diagonal, 647, E-08028-Barcelona, Spain
}

\maketitle

\begin{abstract}

A dynamical systems approach to competition of Saffman-Taylor fingers 
in a Hele-Shaw channel is developed. This is based on the global study of the
phase space structure of the low-dimensional ODE's defined by the 
classes of exact solutions of the problem without surface tension. 
Some simple examples are studied in detail. A general proof of the 
existence of finite-time singularities 
for broad classes of solutions is given. Solutions leading to finite-time
interface pinch-off are also identified. The existence of a continuum of
multifinger fixed points and its dynamical implications are discussed.  
The main conclusion is that exact zero-surface tension solutions taken 
in a global sense as families of trajectories in phase space spanning 
a sufficiently large set of initial conditions, are unphysical 
because the multifinger fixed points are nonhyperbolic, and an unfolding 
of them does not exist within the same class of solutions.
Hyperbolicity (saddle-point structure) of the multifinger fixed points
is argued to be essential to the physically correct 
qualitative description of finger competition.
The restoring
of hyperbolicity by surface tension is discussed as the
key point for a generic Dynamical Solvability Scenario which is
proposed for a general context of interfacial pattern selection.

\end{abstract}


\begin{multicols}{2}

\narrowtext

\section{Introduction}
\label{intro}

The Saffman-Taylor (ST) problem
\cite{Saffman58,Bensimon86,McCloud95,Tanveer00,Casademunt00}
has played a central role for several decades 
as a prototype system in the study of interfacial pattern 
formation\cite{Langer87,Pelce88,Kessler88,Brener91,Pomeau92,Gollub99}, 
particularly concerning the issue of 
pattern selection\cite{Saffman58,McLean81,VandenBroeck83,Kessler85}.
Despite its elongated existence, the problem continues to pose 
new challenges with the focus now on its \it dynamical \rm aspects
\cite{Couder00}. 
In this sense, the ST problem is becoming instrumental once more in 
gaining insights into possibly generic behavior, due to its relative
simplicity in the context of morphologically unstable interfaces in 
nonequilibrium systems.

Full understanding of the analytical mechanisms leading to 
steady state selection by surface tension as a singular perturbation in 
the problem was not completely achieved until 
the late eighties
\cite{Hong86,Hong87,Shraiman86,Combescot86,Combescot88,Tanveer87,Segur91,Benamar91} 
and the resulting scenario, usually referred to as Microscopic Solvability (MS)
\cite{Langer87,Pelce88,Kessler88}, has currently become a paradigm 
for many other systems for instance in free dendritic growth
\cite{Brener91,Pomeau92,Kessler86a}. 
Such solvability analysis, however, is strictly \it static\rm, 
in the sense that it is concerned
with the existence and linear stability of stationary solutions. 
The importance of dynamics in the process of selection was pointed out 
in Refs.\cite{Casademunt91,Casademunt92,Casademunt94} 
where it was argued that the Saffman-Taylor finger solution 
was not the universal attractor of the problem if the displacing
fluid has a non-negligible viscosity. More recently, the traditional 
MS scenario of selection has not been exempt of some controversy in connection 
with the dynamics of the zero surface tension problem
\cite{Mineev97,Casademunt98,Almgren98,Mineev98b,Sarkissian98,Feigenbaum99,Kessler01}.
The singular effects of surface tension on the dynamics have been 
pointed out as a rather
subtle and challenging issue\cite{Tanveer93,Siegel96a,Siegel96b}
and the possibility of some extension of the MS scenario of selection to the
\it dynamics \rm has been suggested 
\cite{Casademunt00,Sarkissian98,Magdaleno98,Magdaleno99}. 
In any case,  
the study of the dynamics of morphologically unstable interfaces in the
context of Laplacian growth or, more generally, of diffusion-limited 
growth of interfaces in nonequilibrium conditions, has been rather 
elusive to analytical treatment due to the highly nonlinear and 
nonlocal character of the equations. Simplified models of Laplacian 
growth in the absence of surface tension have thus been studied in some 
detail\cite{Derrida92,Krug93,Krug97}, and very recently a new 
impulse to the problem of Laplacian growth
without surface tension, with focus on DLA-like self-similar growth,
 is developing after new analytical
insights\cite{Hastings97,Hastings98,Davidovitch99}, leading to
somewhat controversial conclusions\cite{Barra01}. 
For the viscous fingering problem, 
however, the basic question remains as to what extent the 
zero surface tension problem does capture the physics of the fingering 
dynamics. The direct motivation of this question is the existence 
of large classes of explicit time dependent solutions of the zero surface
tension which are nonsingular and span a great variety of possible 
interface morphologies. To what extent these solutions are qualitatively
or quantitatively describing the dynamical behavior or real systems is 
an outstanding, nontrivial question. As we will see, in some cases they 
may evolve to the correct asymptotic solution with the wrong dynamics.
In others, despite the fact that the solutions are unstable to arbitrary 
perturbations, they may accurately describe the time evolution 
of systems 
with nonzero surface tension. The difficulty of the question lies to 
a large extent in the precise way to formulate it.

The present paper expands and elaborates in depth several 
aspects which were first pointed out in Ref.\cite{Magdaleno98}. 
Our central objective is to contribute to elucidate 
the role of surface tension
in the dynamics of finger competition of the ST problem in the channel 
geometry, beyond the obvious role of keeping the interface stable and
free of singularities and the selection of a stationary state. That is 
to elucidate the qualitative differences between the nonsingular 
solutions of the zero surface tension problem and the real problem with a 
possibly small but nonzero surface tension.
Our main contribution is the development of
a new approach to the issue in terms of the ideas and concepts 
of the theory of Dynamical Systems (DS). With this general point of view, 
we will study in detail some specific classes of solutions of the 
zero surface tension problem. As a byproduct, this will provide 
some interesting results
in the context of Laplacian growth concerning the interplay of finger 
widths and screening effects. However the most important issue will 
be the discussion of the physical deficiencies of the nonsingular 
solutions of the zero surface tension 
problem. This will be addressed with the aim at the maximum possible 
generality. 
As we will see, the comparison of the problem with and without surface 
tension is necessarily qualitative in nature, so it is important to 
pose questions in a framework which is at the same time qualitative and 
mathematically precise. Such framework is the theory of Dynamical Systems.
The use of this conceptual tool will help us formulate precise questions 
to which we can give an answer. From the above results and within this 
spirit we will reformulate 
the issue of a possible extension to \it dynamics \rm of the MS scenario of 
steady state selection, and suggest a possible answer to that.

Common understanding of the 
finger competition process (sometimes referred to as finger coalescence) 
leading to the selected steady state is
usually based on qualitative screening arguments \cite{Magdaleno98}. 
In some cases these have been shown to be  
too naive, in accord to the recent findings of stationary 
finger solutions with coexisting unequal fingers \cite{Magdaleno99}, 
as we will discuss in
more detail below. In a first effort to develop precise conceptual tools for
a quantitative and qualitative characterization of fingering dynamics,
a topological approach was developed in Refs.\cite{Casademunt91,Casademunt94} 
to address the nontrivial 
effects of viscosity contrast in the dynamics of finger competition and 
their long time asymptotics. 
The basic insight was to focus on topological
properties of the physical velocity field in the bulk, quantified by the
existence of topological defects. In this paper we will develop a different
global viewpoint, which focuses on topological properties of the 
flow in the (infinite-dimensional) phase space of interface 
configurations, rather than the (two-dimensional) velocity field.

In this paper we will analyze in detail some exact solutions 
of the Saffman-Taylor problem with zero surface tension. Hereinafter 
we will refer to this case as the idealized problem, as opposed to the
regularized one, which will denote the problem with a small but finite
surface tension. Is is known that the idealized ST problem is ill-posed
as an initial-value problem \cite{Tanveer93}. Nevertheless,
one of the crucial facts that makes the ST problem 
attractive from an analytical point of view is the existence of rather 
broad classes of exact time-dependent solutions of the problem without
surface tension 
\cite{Howison85,Howison86,Mineev94,Dawson94,Entov95,Magdaleno00}. 
Some classes of those solutions are known to develop finite-time 
singularities in the form of cusps and are thus of no interest 
to the physics of viscous fingering, but still a remarkably broad 
class of solutions is free of singularities and therefore physically 
acceptable, \it in principle\rm.
The basic question is then what would be the effect of a small but finite 
surface tension to 
those solutions. 
This question was first raised in Ref.\cite{Dai91} where it was shown  
that for some classes of initial conditions, the effect of surface 
tension as a perturbation could be considered as basically regular, 
while  
for other initial conditions the singular character of the perturbation 
showed up dramatically in the dynamics. 
In other configurations, such as for circular geometry, 
surface tension has also been shown to behave as a regular 
perturbation\cite{Dai93}. Indeed, 
in view of the morphological diversity which is included in the
known nonsingular solutions, one may be tempted to believe that, since
such solutions remain smooth for all the time evolution, they should stay
close to the solutions of the regularized problem as $d_0 \rightarrow 0$.
Siegel and Tanveer\cite{Siegel96a,Siegel96b}
have shown that, contrarily to what happens in other
more familiar singular perturbations in physics, in the case of Hele-Shaw
flows that is not the case, and, in general, the idealized and the
regularized
solutions differ from each other at order one time. In the remarkable 
contribution of Refs.\cite{Siegel96a,Siegel96b}, however, 
only simple examples of single-finger evolutions are considered, so the 
extent to which those conclusions can be extrapolated to 
multifinger configurations still requires a careful analysis \cite{Paune01}.
Furthermore, even though the idealized and the regularized solutions 
differ significantly after a time of order unity which is basically 
independent of surface tension, one could still argue that the qualitative
evolution is basically unaffected by surface tension if the finger width 
is not too different from the selected one in the regularized case. 
Therefore, the possibility that some classes of
solutions or some particular dynamic mechanisms are basically insensitive
to surface tension remains open. This question directly motivated the 
work of Ref.\cite{Magdaleno98} which first made use of a dynamical 
systems approach and which we will here pursue further.
The crucial aspect to be exploited is the fact that the integrable 
classes of initial conditions define finite-dimensional invariant 
manifolds of the full (infinite-dimensional) problem, so it makes 
sense to study the resulting low-dimensional dynamical systems 
and compare them with properly defined finite-dimensional subsets 
of the regularized problem. With this analysis we will clarify in what 
precise sense the nonsingular exact solutions of the idealized ST problem are, in general,
unphysical. Once settled the unphysical nature of a broad class of solutions, 
a natural question to address in whether a selection principle is 
associated to the surface tension regularization, which can be understood
as a dynamical generalization of the MS scenario. We will address this point
in the light of our results and discuss how and in what sense such 
dynamical MS could be formulated. 

The rest of the paper is organized as follows. In Sect. \ref{Formulation}
 the equations describing
Hele-Shaw flows in channel geometry are recalled, together with
the conformal mapping formulation used to study the problem. 
The finger competition
phenomenon is described and the relevant quantities to characterize 
finger competition are presented. In Sect. \ref{DS} 
the dynamical systems approach
to the problem is introduced. In Sect. \ref{Minimal} 
the minimal class presented in Ref. \cite{Magdaleno98} is revisited. 
The dynamics of this class, namely,
its phase portrait is studied in detail. The comparison of this phase
portrait with the known topology of the physical problem reveals that
the minimal class dynamics is unphysical. 
The main reason for this unphysical
behavior is the existence of a continuum of fixed points.
In Sect. \ref{Beyond} and \ref{General} various
generalizations of the minimal class are introduced. We pay a special
attention to the perturbation of the minimal class that removes the
continuum of fixed points but keeps the dimensionality of the phase
space unchanged. These solutions
contain the main general features of zero surface tension, and it is
shown that they do not describe in general the correct dynamics for
finite surface tension. The main reason is that the dynamical system
that they define lacks the saddle-point structure of the multifinger fixed
point, necessary
to account for the observed finite surface tension dynamics.
In Sect. \ref{discussion} we discuss the precise role of zero 
surface tension solutions and their relevance to an understanding
of the dynamics of Hele-Shaw flows. In this section 
a Dynamical Solvability Scenario is proposed and discussed as a generalization
of MS theory.
Finally, in Sect. \ref{Concl} we summarize our main results and
conclusions.

\section{Formulation of the problem and characterization of 
finger competition}
\label{Formulation}

Consider a Hele-Shaw cell of width $W$ in the $y$-direction and infinite
length in the $x$-direction, with a small gap $b$ between the plates. 
The fluid flow in this system is effectively two dimensional and the
velocity ${\bf v}$ obeys  Darcy's law
\begin{equation}
\label{Darcy}
{\bf v}=-\frac{b^2}{12\mu}\nabla p
\end{equation}
where $p$ is the fluid pressure and $\mu$ is the viscosity. We define
a velocity potential $\varphi=-\frac{b^2}{12\mu}p$, and assuming that
the fluid is incompressible ($\nabla \cdot {\bf v}=0$) we obtain the
bulk equation to be the Laplace equation
\begin{equation}
\label{Laplace}
\nabla^2\varphi=0.
\end{equation}
This must be supplemented with the two boundary conditions 
\begin{equation}
\label{Gibbs}
\varphi |_\Gamma=d_0\kappa |_\Gamma
\end{equation}
\begin{equation}
\label{continuity}
v_n=\hat{\bf n}\cdot\nabla\varphi
\end{equation}
where $\Gamma$ means that the quantity is evaluated on the interface, $v_n$ is the
normal component of the velocity of the interface, $\kappa$ is the 
curvature, $\hat{\bf n}$ is the unit vector normal to the interface  and
$d_0$ is a dimensionless surface tension defined by $d_0=\frac{\sigma
b^2 \pi^2}{12\mu V_\infty W^2}$, where $V_\infty$ is the fluid velocity
at infinity. Eq. (\ref{Gibbs}) is the Laplace pressure jump
at the interface due to local equilibrium, written in terms of the 
velocity potential. Here we consider the case with the non-viscous 
fluid displacing the viscous one, therefore the pressure in the non-viscous
fluid is constant, and we choose it to be zero. Eq. (\ref{continuity})
is the continuity condition, that states that the interface follows 
the motion of the fluid. 
We assume
periodicity at the sidewalls of the channel. Except for configurations 
symmetric with respect to the center axis of the channel, periodic boundary 
conditions define different dynamics from the more physical case of 
rigid sidewalls (with no-flux through them). Strictly speaking our case 
describes an infinite periodic array of unit channels. We will argue that 
nothing essential is lost with respect to competition in a rigid-wall 
channel, while the analysis is significantly simplified. 

We use conformal mapping techniques to formulate the problem \cite{Bensimon86}.
We define a function $f(\omega,t)$ that conformally maps 
the interior of the unit circle
in the complex plane $\omega$ into the viscous fluid in the physical plane
$z=x+iy$. We assume an infinite channel in the $x$ direction. 
The mapping $f(\omega,t)$ must satisfy $\partial_{\omega}f(\omega,t)
\neq 0$ inside the unit circle, $|\omega|\leq 1$, and moreover, $h(\omega,t)=
f(\omega,t) + \ln \omega$ must be analytic in the interior. 
We define the complex potential as the analytic function 
$\Phi = \varphi + i\psi$, where the harmonic conjugate $\psi$ of
$\varphi$ is the stream function. 
The width of the channel is $W=2\pi$ and the velocity of the fluid 
at infinity is $V_\infty=1$. 
It can be shown that the 
evolution equation for the mapping $f(\omega,t)$ reads \cite{Bensimon86}
\begin{equation}
\label{evold0}
\partial_t f(\omega,t)=\omega \partial_\omega f(\omega,t)
{\rm A}\left[\frac{{\rm Re}(i \partial_\phi \Phi(e^{i\phi},t))}
{| \partial_\phi f(e^{i\phi},t)|^2}\right]
\end{equation}
where ${\rm A}[g]$ is an integral operator that acts on a
real function $g(\phi)$ defined on the unit circle.
${\rm A}[g]$ has the form
\begin{equation}
\label{operadorA}
{\rm A}[g]=\frac{1}{2\pi}\int_0^{2\pi}g(\theta)
\frac{e^{i\theta}+\omega}{e^{i\theta}-\omega }d\theta,
\end{equation}
and on the unit circle $\omega=e^{i\phi}$ it reads
\begin{equation}
\left.{\rm A}[g]\right|_{\omega=e^{i\phi}}=g(\phi)+i{\rm H}_\phi[g]
\end{equation}
where ${\rm H}_\phi[g]$ is the so-called Hilbert transform of $g(\phi)$
defined by
\begin{equation}
{\rm H}_\phi[g]=\frac{1}{2\pi}P\int_0^{2\pi}g(\theta)
\cot(\frac{\phi - \theta}{2})d\theta
\end{equation}
where $P$ stands for the principal value prescription.
The complex potential $\Phi$ satisfies
\begin{equation}
\label{potential}
\Phi(\omega,t)=-\ln \omega + d_0{\rm A}[\kappa]
\end{equation}
where the curvature $\kappa$ given in terms of $f(e^{i\phi},t)$ is
\begin{equation}
\label{curv}
\kappa=-\frac{1}{|\partial_\phi f|}{\rm Im}\left[\frac{\partial_\phi^2 f}
{\partial_\phi f}\right].
\end{equation}
The evolution equation (\ref{evold0}) written on the unit circle 
$\omega=e^{i\phi}$ is 
\begin{equation}
\label{evolcer}
{\rm Re}\{i\partial_{\phi}f(\phi,t) \partial_t f^*(\phi,t)\}=
1 - d_0\partial_\phi {\rm H}_\phi[\kappa]
\end{equation}
where $f(\phi,t)\equiv f(e^{i\phi},t)$.
In the zero surface tension case $d_0=0$ the integro-differential
equation (\ref{evolcer}) reduces to a much simpler
equation, and the evolution of $f(\phi,t)$ for $d_0=0$
is then given by
\begin{equation}
\label{eq:evol}
{\rm Re}\{i\partial_{\phi}f(\phi,t) \partial_t f^*(\phi,t)\}
=1.
\end{equation}  
The direct motivation of the present study is that, despite 
the fact that neglecting surface tension is in principle incorrect from 
a physical standpoint, the $d_0=0$ case can be solved explicitly in 
many cases\cite{Saffman58,Howison86,Dawson94} including solutions 
which, although being unstable, they exhibit a smooth and physically 
acceptable (nonsingular) behavior, 
quite similar to what is observed in experiments 
and simulations of the full problem.

Before proceeding to the description of the general approach and its 
application to specific solutions, let us first introduce some ideas 
and definitions which will be helpful in further discussion. To 
quantify finger competition it is useful to define individual growth 
rates of fingers \cite{Magdaleno98}. In simple situations like those 
considered in the paper, the growth rate of a finger can be simply 
defined (in the reference frame moving with the mean interface position) 
as the peak-to-peak difference of the
stream function between the maximum and the minimum which are adjacent 
to the zero of the stream function located at (or near) the finger tip 
\cite{Casademunt94} (the definition can be generalized to more complicated 
situations).
According to this definition, one
 assigns a  nonzero growth rate to the finger if 
the tip advances at  
a velocity which is larger than the mean interface position. 
Looking at individual growth rates one can easily distinguish two different
stages of the dynamics in the process of finger competition. 
A first stage characterized by the monotonous growth of all individual 
finger growth rates and a second one
dominated by the redistribution of the total growth rate among the fingers. 
We call these two stages \it growth \rm and 
\it competition \rm regimes respectively.  
For a configuration of two different fingers, which is practically 
the only one addressed throughout this paper, during the growth regime
the two fingers develop from small bumps of the initially flat
interface, while the total growth rate $\Delta \psi_T(t)$, defined as
$\Delta \psi_T(t) = \Delta \psi_1(t) + \Delta \psi_2(t)$, grows 
until it reaches a value close to its asymptotic one 
$\Delta \psi_T(\infty)$. 
The decrease of the growth rate of one of the
fingers signals the outcome of the competition regime: there is a
redistribution of flux from one finger to the other one. We also define
the existence of  \it successful \rm competition as the ability to 
completely suppress  the growth rate of one finger. 
A finger
is dynamically suppressed of the competition process when
its growth rate $\Delta \psi$ is reduced to zero. 

\section{Dynamical systems approach}
\label{DS}

The theory of Dynamical Systems is a mathematical 
discipline which studies ordinary differential equations or flows 
(and also difference equations or maps) with stress on geometrical 
and topological properties of solutions \cite{Guckenheiner83}. 
The approach is sometimes
referred to as qualitative theory of differential equations. The
focus is not on the study of individual solutions or trajectories of 
the differential equation, but on global properties of families of 
solutions. This point of view has become very fruitful in searching
for universality in the context of nonlinear phenomena. 

A dynamical systems approach seems thus appropriate to study in a 
mathematically precise way, the qualitative properties of the dynamics
of our problem, and the qualitative differences associated to the
presence or absence of surface tension. One of the important concepts
in dynamical systems theory is that of \it structural stability\rm, which
captures the physically reasonable requirement of robustness of the 
mathematical description to slight changes in the equations. 
Roughly speaking, a system 
is said to be structurally stable if slight perturbations of the 
equations yield a topologically equivalent phase space flow. 
Although the structural stability 'dogma' must be taken with some
caution \cite{Guckenheiner83}, a structurally unstable description 
of a physical problem must be seen in principle as suspect.  
When a dynamical system (DS) depends on a set of parameters, 
the bifurcation set is defined 
as those points in parameter space where it is structurally unstable. In this
case the structural instability at an isolated point in parameter 
space is the property necessary for the system to change its qualitative
behavior. At a bifurcation point, adding perturbations to the equations 
to make the system structurally stable 
is called an \it unfolding \rm \cite{Guckenheiner83}. For dimension 
higher than two, the mathematical definition of structural stability is
usually too stringent. For the purposes of the present discussion and 
most physical applications it is sufficient to consider the notion of 
\it hyperbolicity \rm of fixed points, which in 2 dimensions is directly 
associated to structural stability through the Peixoto theorem \cite{Guckenheiner83}. 
A fixed point is hyperbolic when the 
linearized flow has no marginal directions, that is, all eigenvalues 
of the linearized dynamics are nonzero. We will see that the non-hyperbolicity
of the double-finger fixed point (in general the $n$-equal-finger fixed 
point) and the non-existence of an unfolding of 
it within the known class of solutions is at the heart of the
unphysical nature of this class of solutions.

In the approach to the Saffman-Taylor problem with concepts of DS theory, 
there is, however, an important additional 
difficulty
in the fact that our problem is infinite-dimensional and unbounded. In similar 
spatially extended systems, 
such as described by PDE's, it is customary to project the dynamics onto 
effective low-dimensional dynamical systems based on the so-called 
center manifold reduction theorem\cite{Guckenheiner83}. 
This is possible near the instability 
threshold and, for truly low-dimensional reduction, 
only for strongly confined systems, with a discrete set of modes. 
The ST problem 
however is both unbounded and will operate in general far from
threshold. On the other hand, since the growth is never saturated to
a finite amplitude, any weakly nonlinear analysis is necessarily 
limited to a rather early transient\cite{Alvarez01}. All the 
above techniques are thus of no much use for our purpose of studying 
the strongly nonlinear dynamics of competing fingers in their 
way to the ST stationary solution.  

The basic point that we will exploit here to gain some analytical 
insight into the dynamics of the ST problems as a dynamical system 
is the fact that all exact solutions known explicitly for 
the idealized problem ($d_0=0$) are defined in terms of ODE's 
for a finite number of parameters, and thus define finite-dimensional
DS's in the phase space defined by those parameters \cite{footnote0}. 
The complete ST problem, for any finite $d_0$, defines a 
DS in an infinite dimensional phase space. We will refer to this 
DS as $S^{\infty}(d_0)$. The limit $d_0 \rightarrow 0$ defines a 
limiting DS which we will refer to as $S^{\infty}(0^+)$, which, 
as we will see, is different from $S^{\infty}(0)$.

The conformal mapping $f(\omega,t)$ of the reference unit disk in the 
complex $\omega$-plane into the physical region
occupied by the viscous fluid $z=x+iy=f(\omega)$ has the form 
\begin{equation}
f(\omega,t) = -\ln \omega + h(\omega,t)
\end{equation}
where $h(\omega,t)$ is an analytic function in the whole unit disk,
and therefore has a Taylor expansion 
\begin{equation}
\label{Taylor}
h(\omega,t) = \Sigma_{k=0}^{\infty} a_k(t) \omega^k 
\end{equation}
which is convergent in the whole unit disk.
Inserting Eq.(\ref{Taylor}) into the equation for the mapping 
$f(\omega,t)$ we find an infinite set of equations of the form
\begin{equation}
\label{set}
\dot{a}_k = g_k(a_0,...,a_{k};d_0).
\end{equation}
In the co-moving frame of reference the precise form of the 
infinite set of equations (\ref{set}) is
\begin{eqnarray}
\dot{a}_0=&C_0&\\
\dot{a}_k=&C_k& - kC_0 a_k - \sum_{j=0}^{k-1}j a_j C_{k-j}
\end{eqnarray}
where
\begin{eqnarray}
C_0=&\frac{1}{2\pi}\int_0^{2\pi}\nu(\theta,t)&d\theta\\
C_k=&\frac{1}{\pi}\int_0^{2\pi}\nu(\theta,t)&e^{ik\theta}d\theta
\end{eqnarray}
and
\begin{equation}
\nu(\theta,t)=\left.\frac{{\rm Re}[\omega \partial_\omega h(\omega,t)]
   -d_0 {\rm Re}[\omega \partial_\omega {\rm A}[\kappa](\omega,t)]}
  {|\omega \partial_\omega f(\omega,t)|^2}\right|_{\omega=e^{i\theta}}
\end{equation}

The infinite set of equations (\ref{set}) defines the DS 
$S^{\infty}(d_0)$. In the special case of strictly zero 
surface tension, the DS $S^{\infty}(0)$ can be explicitly 
solved for some classes of initial conditions. These classes
define invariant manifolds of $S^{\infty}(0)$ of \it finite \rm 
dimension. In this context, finding explicit solutions implies 
identifying a specific analytic structure of $h(\omega)$, with 
a finite number of parameters, which is preserved under the 
time evolution. If this condition is fulfilled, then a set of 
ODE's for those parameters can be closed, and defines a certain
DS on a finite-dimension space.
For instance, for $d_0=0$ the truncation of $h(\omega)$ 
into a polynomial form is preserved by the time evolution, so 
Eqs.(\ref{set}) themselves remain a finite set of ODE's. This simple 
case, however, is known to lead to finite-time singularities. 
The evolution is in general not defined after 
some finite time and cannot be considered as sufficiently well 
behaved as a DS's. On the 
other hand, classes of solutions have been reported which are 
smooth (non-singular) for all the time evolution. 
The corresponding conformal mapping takes the general form 
\cite{Howison86,Dawson94}
\begin{equation}
\label{logmap}
h(\omega) = d(t) + \sum_{j=1}^N \gamma_j 
\ln (1 - \alpha_j(t)\omega)
\end{equation}
where $\gamma_j$ are constants of motion with the restriction
$\sum_{j=1}^N \gamma_j=2(1 - \lambda)$ where $\lambda$ is the asymptotic
filling fraction of the channel occupied by fingers. If all $\gamma_j$ are
real the evolution is free of finite-time singularities, and if any
$\gamma_j$ has an imaginary part then finite-time singularities may appear
for some set of initial conditions (see Sect. \ref{cusps}).
Although this form of the mapping contains all orders of the Taylor
expansion of $h(\omega)$, it defines finite dimension invariant
manifolds, since the superposition of logarithmic terms of
Eq.(\ref{logmap}) is preserved under the dynamics, this means that
a closed set of ODE's for the finite number of parameters
$\alpha_j(t)$ can be found. In addition, the region which is physically
meaningful is that in which $|\alpha_j| \le 1$ (including the 
equal sign allows for the limiting case of infinite fingers, and makes 
the phase space compact). The DS defined by Eq.(\ref{logmap}) in the 
$2N$-dimensional hyper-volume will be denoted as $L^{2N}(\{\gamma_j\})$ 

For the sake of discussion throughout this paper it is important to 
have in mind that modifying parameters $\{\gamma_j\}$,
which are constants of motion under the dynamics defined through
Eq.(\ref{eq:evol})
corresponds to varying initial conditions in 
the phase space of $S^{\infty}(0)$, while, from the standpoint of
the finite-dimensional DS's denoted by $L^{2N}(\{\gamma_j\})$ it
corresponds to changing the DS itself, that is, changing the 
ODE's obeyed by the dynamical variables. In this sense,
$\{\gamma_j\}$
label a set of DS's defined on a 2N-hyper-volume $|\alpha_j|\leq1$.

Following Ref. \cite{Magdaleno98}, the key idea is to look for the simplest 
of the DS defined above which contains the three physically relevant 
fixed points, namely, the planar interface (PI), 
the single finger ST solution  (1ST) and the
double Saffman-Taylor finger solution (2ST). We will call 
this minimal DS as $L^2(\lambda,0)$ or simply $L^2(\lambda)$,
since it has only one constant of motion, namely $\lambda$.
In Ref. \cite{Magdaleno98} we proposed to compare the global flow 
properties in phase 
space of such DS with those of a corresponding two-dimensional 
dynamical system defined by the regularized problem. The latter 
was obtained by restricting $S^\infty(d_0)$ to a one-dimensional 
set of initial conditions properly chosen in such a way that 
the invariant manifold of $S^\infty(0)$ which defines 
$L^2(\lambda)$ was tangent to $S^\infty(0^+)$ at the PI fixed point 
\cite{footnote1}.
The resulting DS $S^2(d_0)$ was then 
shown to have a topological structure nonequivalent to that of 
$L^2(\lambda)$. In particular in the limiting case, $S^2(0^+)$  
intersected $L^2(1/2)$ not only at PI but also at the other 
fixed points 1ST and 2ST, and furthermore, at the 
two full trajectories connecting PI with 1ST and 2ST respectively.
The basic conclusion was then that the regularized and the 
idealized problem were intrinsically different.

In this paper we will see that the conclusions drawn from that 
comparison do hold beyond that 'minimal model' analysis. 
The extent and interpretation to which those conclusions 
must be understood will be more precisely stated. We will also 
discuss on the implication of those results on the issue of 
dynamical selection.

\section{The two-finger minimal model}
\label{Minimal}

\subsection{The model}

The simplest exact time-dependent solution of Eq. (\ref{eq:evol})  
containing the three physically relevant fixed points: the planar 
interface (PI), the single Saffman-Taylor (1ST) fixed point and
the double Saffman-Taylor (2ST) fixed point was introduced in 
Ref. \cite{Magdaleno98} and reads
\begin{eqnarray}
\label{eq:map}
	f(\omega,t)=- \ln \omega+d(t)+
	(1 - \lambda) \ln (1 - \alpha(t)\omega)\nonumber\\
	+(1 - \lambda) \ln (1 + \alpha(t)^* \omega)
\end{eqnarray}
where $\lambda$ is a real-valued constant in the interval $[0,1]$, 
$\alpha(t)=\alpha'(t)
+i\alpha''(t)$ and $d(t)$ is real. The relevant phase space for a given
$\lambda$ is the first quadrant of the unit circle in the 
$(\alpha',\alpha'')$ space.  The other three quadrants describe
interface configurations that are equal or symmetrical to 
the interfaces contained in the first quadrant. 
The interface described by  this 
mapping consists generically of two unequal fingers, axisymmetric and without 
overhangs. The axisymmetry  of fingers simplifies the analysis reducing
the number of variables of the problem, but it is important to remark that 
it plays no role in preventing finger competition, as seen in the 
regularized problem with identical symmetry. The case $\alpha'(t)=0$ 
is equivalent to the time-dependent 
ST finger solution \cite{Saffman59}, and $\alpha''(t)=0$ corresponds to the double 
time-dependent ST finger. For $|\alpha(t)|\ll 1$ the interface consists of a
sinusoidal perturbation of the planar interface. 
$d(t)$ affects the overall interface position but does not affect its shape,
and it turns out to be irrelevant for the present discussion.

It is convenient to parameterize the phase space using the variables
$u=1-\alpha''^2$ and $r=(\alpha'^2+\alpha''^2-1)/(\alpha''^2-1)$. Then, 
the new phase space is the square $[0,1]\times[0,1]$. With these 
variables $(u,r)$ the Saffman-Taylor single finger corresponds to the point 
$(0,1)$, the double Saffman-Taylor finger corresponds to $(1,0)$ and the
planar interface to $(1,1)$. 
Substituting the ansatz Eq. (\ref{eq:map}) in Eq. (\ref{eq:evol}), 
we obtain that
the temporal evolution in  the new variables is given by
\begin{eqnarray}
\label{eq:evol1}
\dot{u}=2ru(1-u)\frac{3r-4-gr(1-ru)}{1+gT_g(u,r)}\\
\label{eq:evol2}
\dot{r}=2r(1-r)\frac{3r-2(1+ru)+g(1-ru)(2-r)}{1+gT_g(u,r)}
\end{eqnarray}
where
\begin{eqnarray}
\label{eq:TdeG}
T_g(u,r)=(1-g)(2r+g(2r-1))-\frac{1}{2}(1-g)^2ru\nonumber\\
-gur^2(1+g(ru-3))
\end{eqnarray}
with $g=1-2\lambda={\rm const.}$ The detailed study of the two-dimensional
dynamical system defined by these equations will be the object of the
following section.

\subsection{Study of the dynamical system}
 
The phase portrait of the dynamical system defined by Eqs. (\ref{eq:evol1}-
\ref{eq:TdeG}) with $\lambda=1/2$ is shown in Fig. 1. Its most interesting
feature is the fact that the basin of attraction of the Saffman-Taylor single
finger is not the whole phase space. 
The separatrix between the basin of attraction of the ST finger and the rest
of the flow starts in the planar interface fixed
point and ends in a new fixed point located at $u^*=0$ and 
$r^*=2\lambda /(1+\lambda)$. The flow in the region below the separatrix 
is not attracted to  a single fixed point, but it evolves to a continuum 
of fixed points, the line $r=0$. All these fixed points correspond to 
stationary solutions of the zero surface tension problem consisting of
two unequal fingers advancing with the 
same velocity. The fingers have
different width and tip position, the widths $\lambda_{1,2}$ given by
\begin{equation}
\lambda_{1,2}=\frac{\lambda}{2}\left[1\pm\frac{2}{\pi}{\rm cotg}^{-1}
\sqrt{\frac{u}{1-u}}\right]
\end{equation} 
and the tip separation in the propagation direction $x$ being
\begin{equation}
\Delta_x=(1-\lambda)\ln \frac{1+\sqrt{1-u}}{1-\sqrt{1-u}}.
\end{equation}
The complexity of the phase portrait shown in Fig. 1 is not at all evident
with the original variables $(\alpha',\alpha'')$, as can be seen in Fig. 7a.
Had we not used the parameterization $(u,r)$, the non-trivial structure
of the ST limiting solution would have remained hidden. 
The points along the line
$u=0$ (equivalent to the point $\alpha'=0$ and $\alpha''=1$ in the
original  variables) can be assimilated to  the Saffman-Taylor 
finger because the dominant finger has the asymptotic ST finger 
shape for the corresponding $\lambda$.
All these points differ only in the evolution of the second finger, 
which always evolves 
to a needle (a finger of zero width), but of different length. 
When a trajectory approaches the 
1ST fixed point from the two finger region the second finger does not 
disappear but remains frozen, that is, with $\lim_{t\rightarrow\infty} 
L_S/L_L=0$ where $L_S$ and $L_L$ are the lengths of the short and 
the long finger
respectively. At the other extreme of the $u=0$ line, the point $u=r=0$,
the two fingers satisfy  $\lim_{t\rightarrow\infty}L_S/L_L=1$, but
the distance between the tips goes to infinity. The fixed point $(u^*,r^*)$
corresponds to a new type of asymptotic stationary solution of the zero surface
tension problem, and it consists of two fingers with \it unequal \rm positive 
velocities. Their length ratio satisfies $\lim_{t\rightarrow\infty}L_S/L_L=1/3$
for any $\lambda$.
\begin{figure}[h]
\centerline{{\psfig{figure=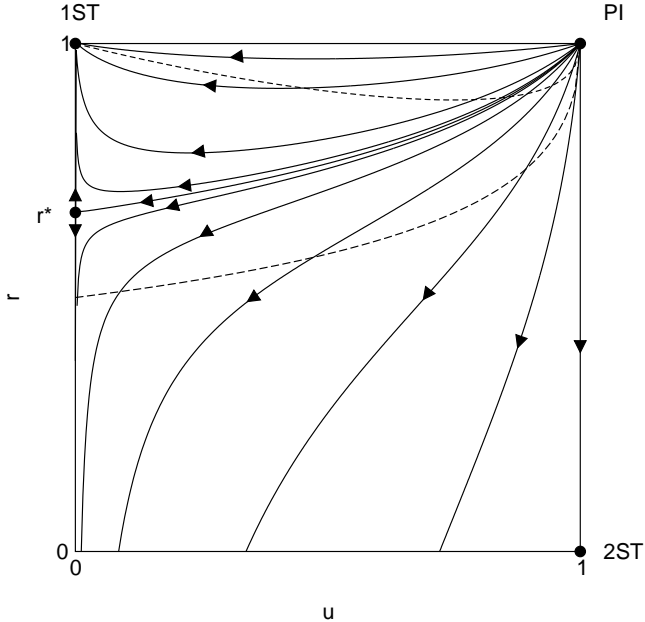,height=9cm,angle=270,clip=t}}}
\caption{Phase portrait of the minimal model with $\lambda=\frac{1}{2}$.
The one-finger (resp. two-finger) region is above (below) the
short-dashed line. For the region above the long-dashed line
the secondary finger has zero growth rate
while for the region below the secondary finger has finite growth rate.}
\end{figure}

One important feature of the phase portrait, regarding the competition 
process, is the position of the line separating the regions where
the small finger has zero (upper) or non-zero (lower) flux (see Fig. 1). 
A crossing
from the region with non-zero flux to the one with zero flux would be the  
characteristic signal of 'successful' competition in the sense 
defined in Section II. But for the physically
relevant value of $\lambda=1/2$, no trajectory crosses 
this line from the lower region or, equivalently, fingers with finite
growth rate keep it finite for all time. Therefore this minimal model does not
exhibit successful competition, for $\lambda=1/2$.

Although $\lambda=1/2$ 
is the physically selected value of the problem regularized
with $d_0 \rightarrow 0$, the study of values of $\lambda\ne 1/2$ may be 
relevant to other purely Laplacian growth problems (without surface tension), 
such as 
DLA or needle-like growth. For fluid fingering problems it may also be 
relevant to situations where the selected finger width differs from $1/2$
due to external perturbations (bubbles at the tip) or anisotropy.
It is thus interesting to extend the analysis to arbitrary $\lambda$.
\begin{figure}[h]
\centerline{{\psfig{figure=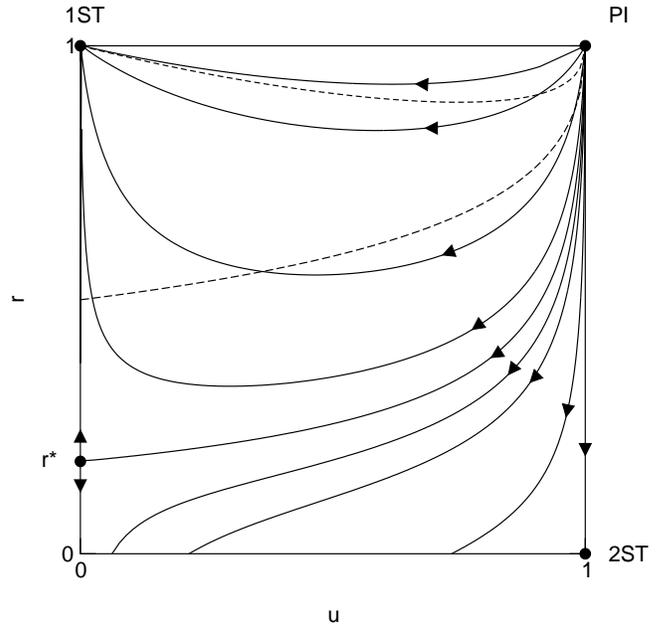,height=9cm,angle=270,clip=t}}}
\caption{Phase portrait of the minimal model with $\lambda=\frac{1}{10}$.
The one-finger (resp. two-finger) region is above (below) the
short-dashed line. For the region above the long-dashed line
the secondary finger has zero growth rate
while for the region below the secondary finger has finite growth rate.
Note that there are trajectories crossing from below the line
separating the zero and finite growth rate regions.}
\end{figure}

The position of the fixed point $r^*$ that separates the 
flux depends monotonically on 
$\lambda$, spanning the whole segment (0,1). For $\lambda > 1/2$,
$r^*$ approaches 1ST. Consequently the basin of attraction of 1ST is
reduced. For $\lambda < 1/2$ the behavior is the opposite: $r^*$ 
departs further from 1ST and 
the basin of attraction of the single finger grows. 
In addition, as $\lambda$ decreases, a critical value $\lambda_c=1/3$
is reached for which $r^*$ crosses the line separating the zero and non-zero
growth rate of the small finger. 
This crossing occurs at the point $u=0$, $r=1/2$. Therefore,
for $\lambda < 1/3$, $r^*$ will be located below that line, and some 
trajectories will cross the line from below: these trajectories exhibit
successful competition, by definition. 
In Fig. 2 it is shown the phase portrait for
$\lambda=1/10$ and trajectories crossing the line from below are depicted.
Despite this fact, such successful competition is rather anecdotic, since
it appears in a region far away from the neighborhood of the double finger,
where the two fingers have a similar flux. The trajectories crossing the
line from below correspond to a second finger with a small growth rate,
thus the amount of flux totally eliminated is relatively small and 
quantitatively not significant
in the competition process compared to the regularized problem. 
In general, what we can  say is that under zero
\begin{figure}[h]
\centerline{{\psfig{figure=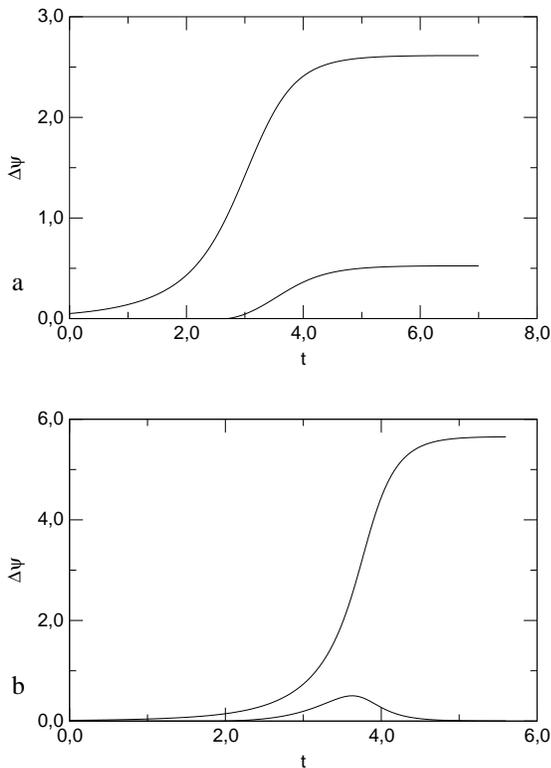,height=11.1cm,clip=t}}}
\caption{Individual growth rates $\Delta \psi_1(t)$ and
$\Delta \psi_2(t)$ of the two fingers for the minimal model
with (a) $\lambda=\frac{1}{2}$ and (b) $\lambda=\frac{1}{10}$.
For the (a) case the trajectory is attracted to
a configuration of two unequal fingers and for
the (b) case the trajectory reaches the Saffman-Taylor finger
with successful competition.
The growth rate of the secondary finger is the curve below.}
\end{figure}
\noindent surface tension dynamics, narrow fingers present 'stronger' competition
than wide fingers. In Fig. 3 we show the evolution of the flux 
for a trajectory ending in a point of the continuum (case with $\lambda=1/2$)
and a trajectory exhibiting successful competition (case with $\lambda=1/10$).

\subsection{Comparison with the regularized dynamics}
\label{comp1}

We are interested in the comparison between the $d_0=0$ dynamics and the 
$d_0 \ne 0$ one. The dynamical system defined by the mapping 
Eq. (\ref{eq:map}) is referred to as $L^2(\lambda)$. From now on
we will restrict the analysis to most relevant case of 
$\lambda=1/2$. 
We introduce the following construction in order to compare the dynamics
with and without surface tension:
consider a one dimensional set of initial conditions ($t=0$) of the 
dynamical system (\ref{eq:evol1}-\ref{eq:TdeG}) surrounding the planar interface (PI) fixed
point $u=1$, $r=1$. An example for this set is simply a quarter of circle 
of small radius centered at $(1,1)$, 
given by $(1-R\cos(\theta), 1-R\sin(\theta))$
with $0\leq \theta \leq \pi/2$ and $R \ll 0$.
 For a fixed $\lambda$, the mapping Eq.
(\ref{eq:map}) applied to this set of initial conditions 
defines a continuous uni-parametric family of interfaces, where $\theta$ 
is the parameter in this example. The evolution up to 
$t \rightarrow \infty$ and backwards to $t \rightarrow -\infty$ 
of the case with $d_0$ finite from this initial uni-parametric set spans 
a compact two-dimensional phase space (referred to as $S^2(d_0)$)
embedded in the infinite dimensional
space of the problem with finite surface tension, $S^\infty(d_0)$. 
It is known from all
experimental and numerical evidence that for finite surface tension the 
subspace $S^2(d_0)$ must contain (at least) three fixed points. 
These three fixed points
are the (unstable) planar interface (PI), the (stable) ST single finger
(1ST') and a saddle fixed point that corresponds to the degenerate double
ST finger (2ST'), where the primes denote the finite surface tension case.
The Saffman-Taylor fixed point (1ST') is known to be the universal attractor
for finite surface tension, so all the trajectories starting in the planar
interface (PI) end up at 1ST'. The saddle fixed point 2ST' has an 
attracting manifold of lower dimension, defined by $\alpha''=0$,
and will govern the dynamics of finger competition. 
We define the space $S^2(0^+)$ as 
the limit of $S^2(d_0)$ for $d_0 \rightarrow 0$.

In the neighborhood of the planar interface, namely the linear 
regime, the surface tension acts as a regular perturbation and 
the regularized problem for small $d_0$ converges regularly to 
the zero surface tension case.
Thus, the manifolds $L^2(1/2)$ and $S^2(0^+)$ must be tangent
at the PI fixed point $u=1$, $r=1$.
Moreover, from selection theory we know that 
1ST'$\rightarrow$1ST and 2ST'$\rightarrow$2ST in the limit 
$d_0 \rightarrow 0$, provided that $\lambda=1/2$ is set for the 
zero surface tension manifold. Therefore,
$L^2(1/2)$ and $S^2(0^+)$
must be tangent at 1ST and 2ST.

We have shown that the fixed points of the regularized space $S^2(0^+)$
are contained in the zero surface tension one $L^2(1/2)$,
but the $L^2(1/2)$
contains additional fixed points that are not part of the 
regularized problem $S^2(0^+)$. As we have seen in the previous 
section, these additional fixed points are the continuum of
two-finger steady solutions.
Therefore, it is important to stress that the flow for the two-dimensional
subspace $S^2(0^+)$ defined above is \it not \rm topologically 
equivalent to the flow $L^2(1/2)$
of the minimal model. More precisely, the $d_0=0$ dynamical system
contains a new fixed point that separates the phase space in
two basic regions: the set of points ending up at 1ST and the set not 
ending up at 1ST. Despite the fact that we do not know explicitly 
the trajectories of the regularized 
problem in the phase flow $S^2(0^+)$, we can conclude that the 
two cases are fundamentally different.

We have chosen the minimal model  with the requirement of including the
double degenerate ST finger fixed point  because it is essential to
the competition process. 
In fact, from a physical point of view it is clear that 
the 2ST fixed point must govern the
crossover from the growth regime to the competition one and its
saddle point structure is necessary to the physical phenomenon of
finger competition, in the sense that it has an attracting manifold
(containing the PI fixed point) associated to the growth regime, and
an unstable manifold associated to the competition regime. 
However, the
zero surface tension dynamics fails to reproduce the saddle-point 
structure necessary to describe the crossover from growth to competition.
This failure is not a particularity of the minimal model but a generic
property of the zero surface tension problem, as we will see 
in the following sections.

The two-dimensional dynamical system (\ref{eq:evol1},\ref{eq:evol2},\ref{eq:TdeG}) 
defines a structurally unstable flow, in application of the Peixoto
theorem\cite{Guckenheiner83}, due
to the existence of the continuum of fixed points at the line $r=0$. 
This structural instability is clearly related to the unphysical
behavior of the minimal model, and its effects will be felt
by solutions formally close to (\ref{eq:map}), as it will be shown in
the following section. According to the theory of dynamical systems,
the introduction of an arbitrarily small perturbation to the
dynamical system equations (\ref{eq:evol1},\ref{eq:evol2},\ref{eq:TdeG}) 
could act as an 
unfolding of the problem, suppressing the continuum of fixed points and 
replacing the non-hyperbolic 2-equal-finger fixed point by an isolated 
one with a saddle-point (hyperbolic) structure.
In this way, the unfolding of the phase portrait defined by the minimal class
(\ref{eq:map}) would have the same topology as the physical problem. Therefore, 
the natural step is to introduce a perturbation  to the mapping (\ref{eq:map})
that should unfold the phase portrait and still be solvable. The next section is devoted to
such perturbed ansatz.
\vspace{1cm}

\section{Extension within two dimensions: searching for 
an unfolding}
\label{Beyond}

\subsection{Modified minimal model}

A possible modification of the ansatz (\ref{eq:map}) 
which is solvable and preserves the two-dimensionality of the  phase space is the following:
\begin{eqnarray}
\label{fepsilon}
	f(\omega,t)=- \ln \omega + d(t) +
	(1 - \lambda + i\epsilon) \ln (1 - \alpha(t)\omega)\nonumber\\
	+(1 - \lambda - i\epsilon) \ln (1 + \alpha(t)^* \omega)
\end{eqnarray}
where $\epsilon$ is a real positive and is a constant of motion. 
If $\epsilon$ is set to zero then the minimal model Eq. (\ref{eq:map})
is retrieved.
Solutions of this type have been studied before for instance in 
Refs.\cite{Mineev94,Mineev99,Feigenbaum99}.
This mapping describes generically two unequal axisymmetric fingers,
with the symmetry axis located in fixed channel positions separated
a distance $\pi$, half the channel width. The main morphological difference 
between the interface described by the minimal class and the interface
obtained from Eq. (\ref{fepsilon}) is that the interface of the modified 
model may present overhangs. 
This can be understood from a geometrical
point of view in the following manner: well developed fingers are
separated from each other by fjords of the viscous phase, and the
width and orientation of these fjords is determined by the constant 
term $1 - \lambda + i\epsilon$. If the constant term is real, 
i.e. $\epsilon=0$, the centerline of the fjords is parallel to the
channel walls and the fingers do not present overhangs, but if the
constant term has an imaginary part ($\epsilon \neq 0$) then 
the fjords form a finite angle with the walls \cite{Dawson94}. As a result of the
inclination of the fjords the fingers may present overhangs.
An example of these solutions is shown in Fig.4, with a series of snapshots 
of the corresponding time evolution. 
The class of solutions Eq.(\ref{fepsilon})
contains also the single finger Saffman-Taylor
solution ($\alpha'=0$) but, remarkably enough,
the introduction of a finite $\epsilon$ has
removed the double Saffman-Taylor finger solution.
\begin{figure}[h]
\centerline{{\psfig{figure=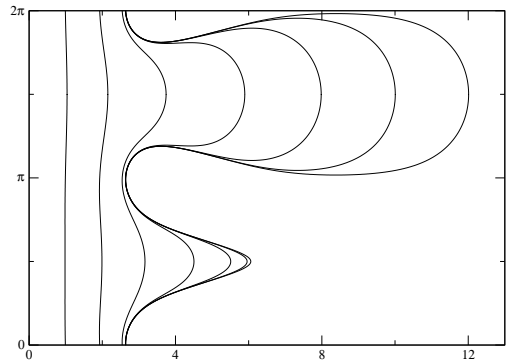,height=5.5cm,angle=270,clip=t}}}
\caption{Time evolution of a configuration with $\lambda=1/2$ and $\epsilon=0.1$.}
\end{figure}
As usual, 
the constant $\lambda$ is the asymptotic width of the advancing finger. 
The natural phase space in this case is the unit circle, 
$|\alpha|\leq 1$, but we will
restrict the study to $\alpha' \geq 0$ because 
the $\alpha' \leq 0$ region can be obtained by a $\pi$ rotation 
of the $\alpha' \geq 0$ region.
Physically, this rotation or the replacement $\alpha \rightarrow - \alpha$ 
corresponds to a shift of the interface by an amount $\pi$ (half
the channel width) in the $y$ direction.
Thus, the semi-circle $\alpha' \leq 0$ contains the
same interfacial configurations and dynamics than the $\alpha' \geq 0$
region after a trivial transformation.
For the minimal model the zeros $\omega_0$ of $\partial_\omega f(\omega,t)$
laid outside the unit circle, but for the modified minimal model 
Eq. (\ref{fepsilon}) the situation is different. For $|\alpha|< 1$
a zero of $\partial_\omega f(\omega,t)$ can be inside the unit circle.
The position of the zeros in this case is
\begin{equation}
\label{pzero}
\omega_0=\frac{-i(\lambda \alpha'' - \epsilon \alpha')\pm
\sqrt{(2\lambda - 1)|\alpha|^2 - (\lambda \alpha'' - \epsilon \alpha')^2}}
{(2\lambda - 1)|\alpha|^2}
\end{equation}
for $\lambda \neq 1/2$ where $\alpha'={\rm Re}\: \alpha$ and 
$\alpha''={\rm Im}\: \alpha$. For the particular value $\lambda=1/2$
the position of the zero is 
\begin{equation}
\label{zeromig}
\omega_0=\frac{1}{2i(\lambda \alpha'' - \epsilon \alpha')}.
\end{equation}
It can be shown that for any $\lambda$ and 
$\epsilon \neq 0$ a $\omega_0$ can be found such that 
$|\omega_0|<1$ for some $|\alpha|<1$.
For instance, with $\lambda=1/2$ the curve $|\omega_0(\alpha)|=1$
is the line $\alpha''=-1+2\epsilon\alpha'$, which clearly intersects
the unit circle $|\alpha|=1$, enclosing a region where $|\omega_0|<1$. 
As a consequence of the presence of a zero inside the unit circle
the parameter space $|\alpha|=1$ 
contains unphysical regions, where the mapping 
Eq. (\ref{fepsilon}) does not describe physically acceptable situations,
with self-intersection of the interface associated to the fact that 
the mapping is not a single valued 
function. One of these regions is defined by the existence of 
a zero $\omega_0$ of $\partial_\omega f(\omega,t)$ inside 
the unit circle. In this region of phase space the interface crosses itself 
at one point, describing a single loop (see an example in Fig.5). 
Most remarkably a second unphysical 
region containing interfaces with two intersections 
cannot be so 
\begin{figure}[h]
\centerline{{\psfig{figure=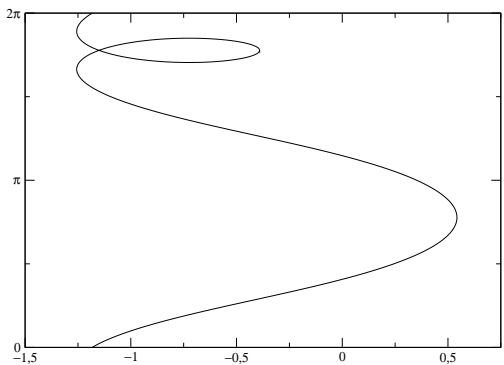,height=5.5cm,angle=270,clip=t}}}
\caption{Interface with one crossing, with one zero inside the unit circle.}
\end{figure}
\noindent easily detected since, in this case, the zeros of 
$\partial_\omega f(\omega,t)$ lay outside the unit circle.
Zero surface tension solutions displaying this feature were 
also reported in Ref. \cite{Baker95}. 
Fig. 6 shows a configuration with this double crossing.
\begin{figure}[h]
\centerline{{\psfig{figure=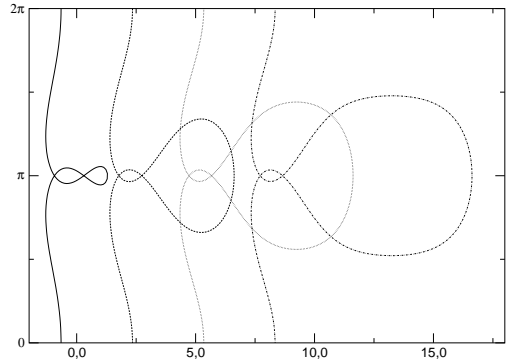,height=5.5cm,angle=270,clip=t}}}
\caption{Time evolution of a configuration with a double crossing
of the interface, with $\lambda=\frac{1}{2}$ and
$\epsilon=\frac{1}{2}$. The solid line corresponds
to $t=0$ with $\alpha=0.85+i0.4$
and the short-long dashed line to $t=3.0$. (The curves
are plotted with its mean $x$ position shifted arbitrarily for better visualization).}
\end{figure}
Substituting the mapping Eq. (\ref{fepsilon}) in Eq. (\ref{eq:evol}) it can be
checked that this ansatz is a solution and that $\epsilon$ is a 
constant preserved by the dynamics. The system of differential
equations resulting from the substitution takes the form
\begin{eqnarray}
\label{sistema-eqsa}
\dot{d} + 4\, {\rm Im} [\alpha (1 - \gamma)]\{ \dot{d}\: {\rm Im}[\alpha] + 
{\rm Im}[\gamma \dot{\alpha}]\} + |\alpha|^2(2\lambda - 1) \times
\nonumber \\ 
\{ \dot{d} |\alpha|^2 + 
2\, {\rm Re}[\gamma \alpha^* \dot{\alpha}]\} = 1 + |\alpha|^4 + 4\, {\rm Im} [\alpha]^2\\
2\{\dot{d}\: {\rm Im} [\alpha] + {\rm Im} [\gamma \dot{\alpha}] \} + 
2\dot{d}\: {\rm Im} [\alpha(1 - \gamma)] + \nonumber \\
2\, {\rm Im} [\alpha(1 - \gamma)] \{ \dot{d} |\alpha|^2 + 
{\rm Re} [\gamma \alpha^* \dot{\alpha} ] \} + \nonumber \\
\label{sistema-eqsb} 2(2\lambda - 1)
|\alpha|^2 \{ \dot{d}\: {\rm Im}[\alpha]  + {\rm Im} [\gamma \dot{\alpha}] \} =
4(1 + |\alpha|^2)\,{\rm Im}[\alpha] \\
\label{sistema-eqsc}
2\lambda \dot{d} |\alpha|^2 + 2\, {\rm Re} [\gamma \alpha^* \dot{\alpha}] = 2 |\alpha|^2
\end{eqnarray}
where the time-dependence of $\alpha(t)$ and $d(t)$ has been dropped for
sake of clarity and $\gamma=1 - \lambda + i\epsilon$. 
Eqs.(\ref{sistema-eqsa}-\ref{sistema-eqsc}) can be integrated explicitely and the corresponding 
solutions for the variables 
$d(t)$ and $\alpha(t)=\alpha'(t)+i\alpha''(t)$ take the form 
\begin{eqnarray}
\label{sistema}
	\beta = d(t) - \ln \alpha(t) + (1 - \lambda - i\epsilon)
	\ln (1-|\alpha(t)|^2) \nonumber \\+(1 - \lambda + i\epsilon)
	\ln (1 + \alpha(t)^{2}) \\
\label{sistema1}
	t + C = \lambda d(t) + (1 - \lambda)\ln |\alpha(t)| -
	\epsilon \arctan \frac{\alpha''(t)}{\alpha'(t)}
\end{eqnarray}
where $C$ is a real-valued constant and $\beta$ is a complex-valued constant. 
Notice that there is no apparent indication of the pathological situations 
described above in the form the explicit solutions above. 
The study of the dynamical systems defined by Eqs. (\ref{sistema},\ref{sistema1}) is the 
object of the next section.
 
\subsection{Study of the dynamical system}
\label{study_ep}

The addition of an imaginary part $i \epsilon$ to the constant $(1-\lambda)$ 
modifies dramatically the phase portrait of the minimal model, 
as can be seen in Fig.7, where the phase portraits for $\epsilon=0$
(now in the variables $\alpha', \alpha''$) and $\epsilon=0.1$ are depicted.
The phase portrait of the modified minimal model is 
qualitatively 
different from the $\epsilon=0$ one, as a direct consequence of the 
structural instability of the latter case \cite{Guckenheiner83},
which implies that an arbitrary perturbation of the equations 
yields a flow which is not topologically equivalent (notice that a 
perturbation of an initial condition of the infinite-dimensional space 
of interface configurations is represented here as a perturbation of 
the equations themselves, 
that is, a displacement in the space of dynamical systems.)
One could have expected that the introduction of this $\epsilon$ would 
have provided an unfolding of the phase portrait of the minimal model into 
a structurally stable one (within the integrable class of mappings), 
hopefully with the saddle-point connection 
between the unstable and the stable fixed points, as corresponds to the
the physical case
with $d_0 \neq 0$. The phase portrait of the regularized flow 
(which is obviously the natural unfolding) 
would be similar to that of $\epsilon=0$ in Fig.7a, except that all 
trajectories other than the line $\alpha''=0$ would end up symmetrically to
the upper ST fixed point or the lower one. Notice that in this 
representation, 
the 1ST fixed point has been split in two 1ST(R) and 1ST(L), corresponding 
to whether the right or the left finger approaches the single finger 
attractor. 
\begin{figure}[h]
\centerline{{\psfig{figure=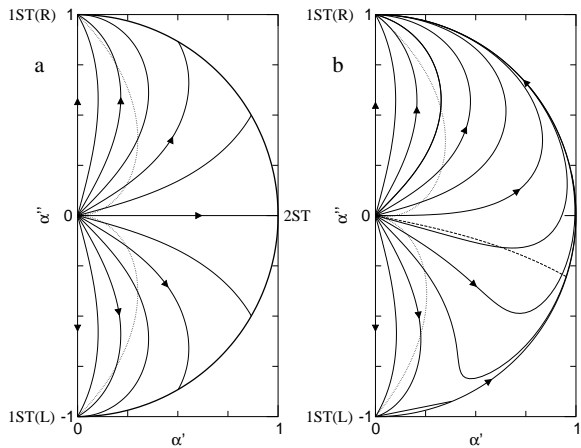,height=7cm,angle=270,clip=t}}}
\caption{Phase portrait of the minimal model and the
modified minimal model.
$\lambda=\frac{1}{2}$ for both plots,
the regions to the right of the dotted lines
correspond to two-finger configurations
(a) $\epsilon=0$; note the continuum of fixed
points (marked with a thick line) on $|\alpha|=1$.
(b) $\epsilon=0.1$; the straight line in the lower left corner is a line of finite-time
singularities and the two fingers have equal length on the dashed line.}
\end{figure}
These two solutions correspond to having the ST finger 
located at two different positions (the symmetry axes of the fingers), owing 
to the translational invariance associated to the periodic boundary 
conditions. Since a $\pi$-shift in the transversal $y$-direction 
must yield an equivalent configuration, 
the identification of any point in the semicircle with the diametrically 
opposed which has reduced the actual phase space in half, implies also that 
the 1ST(R) and 1ST(L) must be topologically identified as the same point.
Approaching one or the other thus means approaching from the left or 
from the right. With this identification the contact with the problem 
with rigid boundaries is clearer, since there, the attractor is clearly 
unique but the flow must also be symmetrically split into two parts, 
corresponding to whether the left or the right finger wins, owing to
the symmetry of the system under parity (see a more detailed discussion in
section VI.D).
Unfortunately, we must conclude that 
the resulting phase portrait for the modified minimal model 
does not provide the correct unfolding. This is particularly remarkable 
if one takes into account that, 
in two-dimensional systems, structurally stable dynamical 
systems are dense \cite{Guckenheiner83}. On the contrary, the perturbed
equations contain finite-time singularities and, although they remove 
the continuum of double-finger fixed points, they also miss the equal-finger 
fixed point, which is an essential ingredient of the regularized flow.

In Fig.8 we plot the phase portrait for $\epsilon=0.5$ and the different 
regions of phase space. For any other 
$\epsilon$ the flow is topologically equivalent but the shape and size 
of the different regions varies smoothly. The line of finite-time singularities
collapses towards the lower fixed point 1ST(L) 
in the limit $\epsilon \rightarrow 0$ 
as shown in Fig.7b. In the absence of the 2ST fixed point, the splitting of 
flow is made possible by the existence of the line of finite-time 
\begin{figure}[t]
\centerline{{\psfig{figure=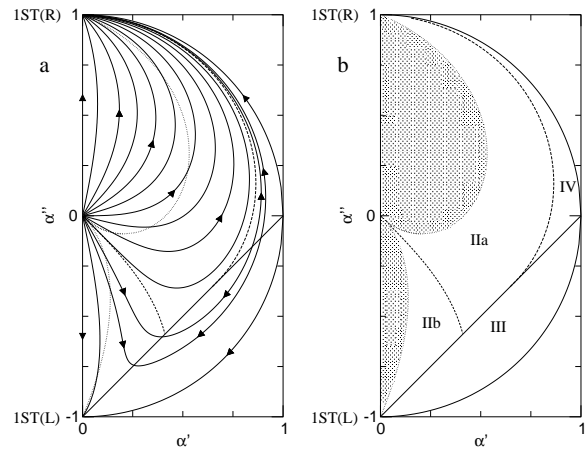,height=7cm,angle=270,clip=t}}}
\caption{(a) Phase portrait of the modified minimal model with $\lambda=1/2$
and $\epsilon=1/2$. (b) Plot of different regions of phase space of case (a). 
The grey regions correspond to single finger interfaces and the other regions to
two finger interfaces. Regions IIa and IIb differ in which of the two fingers is larger.
Regions III and IV are unphysical regions described in the text. The straight boundary 
of region III is a line of cusp singularities.}
\end{figure}
\noindent singularities.
Instead of a separatrix between the respective basins of attraction of
1ST(R) and 1ST(L), there is an
 intermediate, non-zero measure region, connected to 
the PI fixed point,
whose evolution ends up at that singularity line, defined by the condition 
$|\omega_0| = 1$. Similarly to the finite-time singularities occurring for
polynomial mappings, this line is reached in a finite 
time and is associated to the formation of a cusp at the interface.
The evolution
is not defined after that time.
The flow in the region below the singularity line (region III of Fig.8b),
defined by $|\omega_0| < 1$, 
is actually well defined although it describes evolution of unphysical 
interfaces which 
intersect themselves forming a loop, 
(see Fig.5). Their evolution originates and ends at different points of 
the singularity line.
The region IV has double 
crossings of the interface (see Fig.6) and 
also originates at the singularity line
but, remarkably enough, it evolves asymptotically towards the ST finger 
despite their unphysical double crossing at the tail of the finger. 
This double-crossing is removed in a finite time in some subregion of IV 
and it remains up to infinite time in another subregion. Incidentally,
this clearly illustrates how dangerous it may be to infer a physically 
correct dynamics from the fact that the interface evolves asymptotically 
towards a single ST finger, 
and that zero surface
tension solutions must be dealt with extreme care since smooth and
apparently physical interfaces may contain elements that yield them physically
unacceptable when the time evolution is considered either forward or 
backward.
\begin{figure}
\centerline{{\psfig{figure=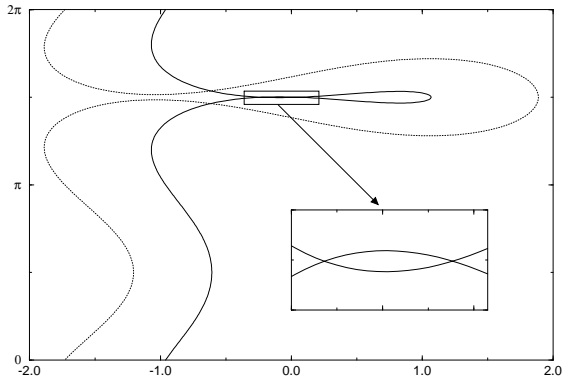,height=5.5cm,angle=270,clip=t}}}
\caption{Time evolution of a configuration with a double crossing
of the interface, showing the dynamical removal of the crossing.
In this case, $\lambda=\frac{1}{2}$ and
$\epsilon=\frac{1}{2}$. The initial interface with $\alpha=0.865+i0.2$ is the solid line,
and the dotted line is the interface at $t=0.5$. The mean $x$ position of the interface
is shifted for better visualization. Time reversal of this evolution, corresponding
to stably stratified Hele-Shaw flow, defines a finite-time interface pinch-off.}
\end{figure}
The double-crossing removal in some of the above solutions 
has some implications in the general study of 
topological singularities associated to interface pinch-off in 
fluid systems (for a recent review see Ref.\cite{Eggers97} and, in the context 
of Hele-Shaw flows, see for instance Ref.\cite{Goldstein98}). 
Consider the stable Saffman-Taylor
problem, in which the viscous fluid displaces the inviscid one. The planar
interface is stable in this case and is the attractor of the dynamics. The
conformal mapping obeys then an equation formally equivalent to 
Eq.(\ref{eq:evol}),
that applies to the unstable Hele-Shaw flow, with the only difference 
that time is reversed, 
$t$ is substituted by $-t$, in Eq.(\ref{eq:evol}).
Therefore, the dynamics of the stable case is obtained 
from the unstable one 
simply by a time reversal. 
As a consequence, the double-crossing removal 
we observed in the original problem encompasses a 
prediction of a finite-time interface pinch-off in the stable 
configuration of the problem, for some class of initial conditions. 
A similar pinch-off phenomenon for zero surface tension dynamics was
detected numerically by Baker, Siegel and Tanveer \cite{Baker95} 
for other types of
mapping singularities. Our result provides a very simple example of 
exactly solvable finite-time pinch-off. Notice that there is no singularity 
of the interface shape or velocity at the interface contact, so one could
presume that surface tension may not affect significantly the phenomenon
in this case, 
although this is an open question yet.

The evolution of a trajectory ending up in the cusp line cannot be 
continued beyond the impact time $t_0$ of the zero $\omega_0$ 
with the unit circle $|\omega| =1$ because the cusp line attracts the 
flow from both sides, but the flow is indeed well defined in the interior
of region III, where we have $\frac{d|\alpha|}{dt} < 0$ in opposition to
the physical region where $\frac{d|\alpha|}{dt} > 0$, so, in a sense, 
the temporal evolution inside the singularity region is reversed.
As a matter of fact, the graph $\alpha''(\alpha')$ can be obtained and is 
smooth in all regions of phase space. 
Defining $\alpha=r e^{i\theta}$, its
substitution in Eq. (\ref{sistema}) yields after some algebra
\begin{equation}
\label{traj}
\frac{d\theta}{dr}=\frac{4r\cos \theta}{1-r^2}
\frac{(1-\lambda)(1-r^2)\sin \theta + \epsilon(1+r^2)\cos \theta}
{1+(2\lambda - 1)r^4 + 2\lambda r^2\cos 2\theta + 2\epsilon r^2 \sin 2\theta}
\end{equation}
and from this expression the trajectory can be obtained also
in region III. The fact that the modified minimal model does not
yield an unfolding of the minimal one is more deeply stressed 
by the fact that the field of directions defined by the above graph,
even removing the singularities through a proper time reparametrization and 
time reversal in region III, is still a structurally unstable flow.

It is well known that the zero-surface tension problem is extremely sensitive
to initial conditions: given a zero-surface tension solution at $t=0$,
another one can be found which is as close as desired 
to the interface of the first solution, and
the evolution of the two solutions be completely different in general.
This is a consequence of the ill-posedness of the initial-value problem 
\cite{Tanveer93}, which is most clearly manifest in polynomial mappings, 
which may arbitrarily approximate any initial condition, possibly one 
which does not develop singularities, but themselves always develop 
cusps. However, it is illustrative to see several striking examples of
sensitivity to initial conditions within the class of logarithmic mappings 
which are much less predictive \it a priori \rm.

\it Example 1 \rm. Consider two initial conditions 
$(\alpha_1',\alpha_1'')$ and $(\alpha_2', \alpha_2'')$ 
close to the PI fixed point, with $|\alpha_1|,|\alpha_2|\ll 1$, which 
differ only in nonlinear orders of their mode amplitudes \cite{modes}.
\begin{figure}[h]
\centerline{{\psfig{figure=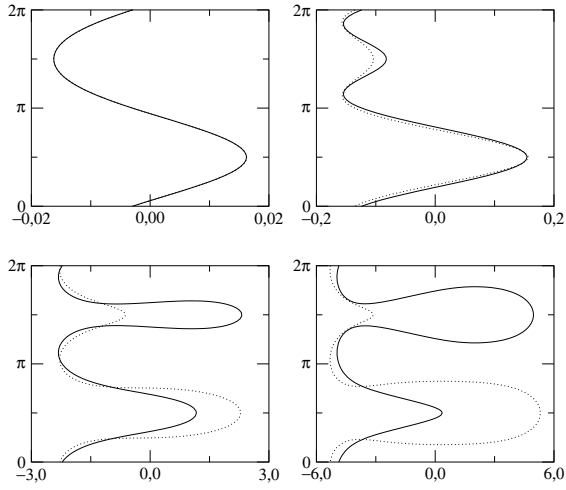,height=7.25cm,angle=270,clip=t}}}
\caption{Evolution of two interfaces initially equal to linear order (see text),
with $\lambda=1/3$ and $\epsilon=0.1$. $\alpha(0)=0.04619398-i0.01913417$ for
the solid line and $\alpha(0)=-0.04619398-i0.00527598$ for the dashed line. 
Upper left plot $t=0$, upper right plot $t=2.0$, lower left plot $t=4.0$ and
lower right plot $t=6.0$.}
\end{figure}
One can easily choose $(\alpha_2',\alpha_2'')$ 
(with $\alpha_1'\alpha_2' < 0$, that is, considering not 
only the semi-circle $\alpha' > 0$ but the whole unit circle)  
such that the time evolution will be completely 
different from the evolution of the original initial condition, even 
though the two initial conditions where equivalent to linear order.
In Fig.10 we show an explicit example.
While the two initial conditions for the interface configuration
cannot be distinguished in the scale of the plot, the final outcome 
is dramatically different. One of the evolutions is an example of successful 
competition, where the finger in the initial condition is eventually approaching
the ST solution, with a small
secondary finger (not present in the initial condition) which is generated 
but screened out by the leading one. The other evolution is quite surprising 
since the secondary finger grows to the point of taking over and winning 
the competition.  
 
\it Example 2 \rm. A similar situation is found if one compares two 
initial conditions equivalent to linear order up to a parity transformation.
Pairs of initial conditions of this type, with the same values of 
$\lambda$ and $\epsilon$, 
 can easily be found within the 
same semicircular phase space, and since the dynamics is indeed symmetric 
under mirror reflection, one should not expect, in principle, a very 
different 
behavior, even though such points are not close to each other in phase space. 
Fig.11 shows an example in 
which one of the evolutions is smooth, with a leading finger and a small 
one being generated, and the other generates a cusp in finite time. As 
in the first example, no signature of the different fate of the system 
could apparently be seen in the initial conditions. In both cases the 
extremely small differences associated to higher orders in the mode 
amplitudes have thus been crucial. 
The sensitivity to initial conditions 
of these examples is more striking for decreasing values of 
$\epsilon$, since the time in which the two evolutions stay close to 
each other increases as $O(- \ln \epsilon)$. For instance,
given an initial condition $\alpha_0$
close to PI, the difference between the $\epsilon=0$ interface and the
$\epsilon \rightarrow 0$ one  will remain of $O(\epsilon)$
for a time of $O(-\ln \epsilon)$.  Later on in the evolution the differences
between the two interfaces will be of $O(1)$: the asymptotic
shape of the $\epsilon=0$ case
will be two unequal fingers while the shape of the $\epsilon \rightarrow 0$
will be a single Saffman-Taylor finger.
Similarly, for two initial conditions symmetrical to linear order such as in 
Example 2, with 
$\epsilon \rightarrow 0$, the differences between their interfaces
will remain symmetric to $O(\epsilon)$ for a time of $O(-\ln \epsilon)$,
but later they will lose symmetry and finally both will end up at the
same fixed point, say the right one, even though one of the two 
evolutions has been favoring the other one, say the left one, for a 
long time (up to well developed fingers).
\begin{figure}
\centerline{{\psfig{figure=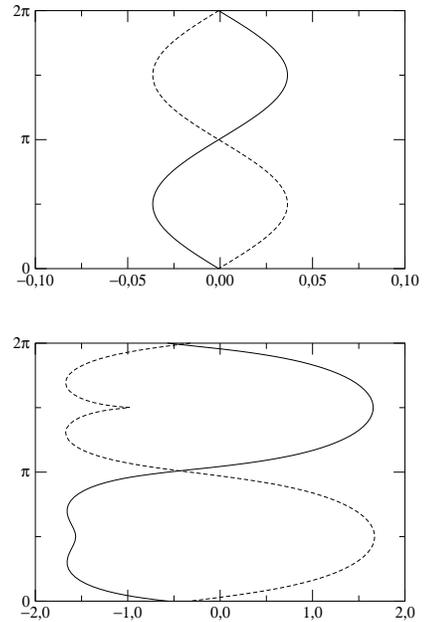,height=9cm,clip=t}}}
\caption{Evolution of two interfaces
symmetric to linear order (see text), with $\lambda= \frac{1}{2}$,
$\epsilon=0.1$, $\alpha(0)=0.02724+i0.03104$ for the solid line and 
$\alpha(0)=0.02724-i0.04193$ for the dashed line. The unpper plot corresponds to
$t=0$ and the lower to $t=4.19$, when a cusp develops.}
\end{figure}

\it Example 3 \rm. In Fig.12 we illustrate the effect of changing $\epsilon$ 
in initial conditions which are equivalent to linear order. Notice that 
in one case a cusp is generated at the secondary finger. In others 
the small fingers rapidly overcomes the large one, while for the smallest 
$\epsilon$ the initial finger seems to lead the competition. Remarkably, 
in this last case the smaller finger will also take over after a much 
longer time.
\begin{figure}[h]
\centerline{{\psfig{figure=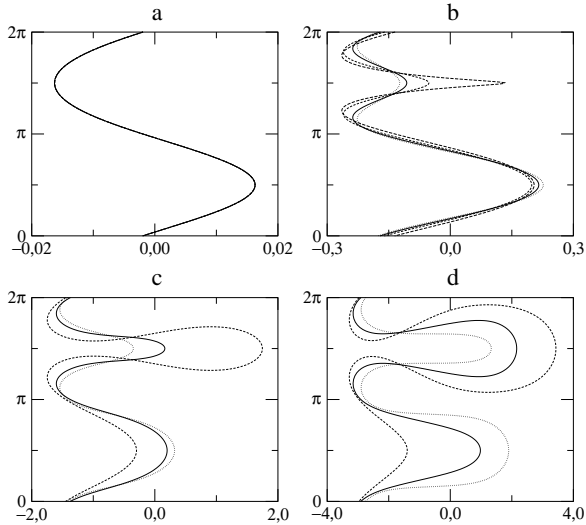,height=7.25cm,angle=270,clip=t}}}
\caption{Evolution of four interfaces equal to linear order,
with $\lambda= \frac{1}{2}$. $\epsilon=0.01$ and $\alpha(0)=0.04671-i0.01291$ for
the dotted line, the solid line is $\epsilon=0.1$ and $\alpha(0)=0.04619-i0.01913$,
the short-dashed line is $\epsilon=0.3$ and $\alpha(0)=0.04260-i0.03137$ and the long-dashed
is  $\epsilon=0.6$ and $\alpha(0)=0.03472-i0.04345$.
(a) $t=0$, (b) $t=2.4$, (c) $t=3.5$, (d) $t=5.0$.
Note that at $t=2.4$
the $\epsilon=0.6$ interface develops a cusp.}
\end{figure}

All the above examples have been chosen to emphasize the caution that 
is required when trying to use exact solutions to approximate the 
dynamics of the problem. A direct comparison of these solutions with 
numerical integration for very small surface tension would be required 
in order to make a more quantitative assessment of the issue. This 
will be presented elsewhere \cite{Paune01}. In any case, it must also be 
stated that the class of logarithmic solutions does provide also 
qualitatively correct evolutions, not only of single finger configurations 
as stated in \cite{Magdaleno98}, but also with two-finger configurations 
showing successful competition. An example of this is plotted in Fig.4.
Starting from the planar interface, during the linear regime
a bump starts to grow, followed generically by a second bump as
the evolution enters the non-linear regime. The two fingers keep on
growing for some time, until
later on in the evolution, one of the fingers is dynamically eliminated of
the competition process and the other finger approaches asymptotically the
ST finger solution. This general scenario is illustrated in figure 13a, where
the individual growth rates of the two fingers $\Delta \psi_1$ and
$\Delta \psi_2$ are plotted versus time, for two different initial
conditions.

For other initial conditions as generic as the previous one, however,
the following phenomenon is observed: the small finger (with initially zero
flux) of a configuration with two significantly unequal 
fingers increases its flux
while the flux of the large finger decreases, until the flux 
\begin{figure}[h]
\centerline{{\psfig{figure=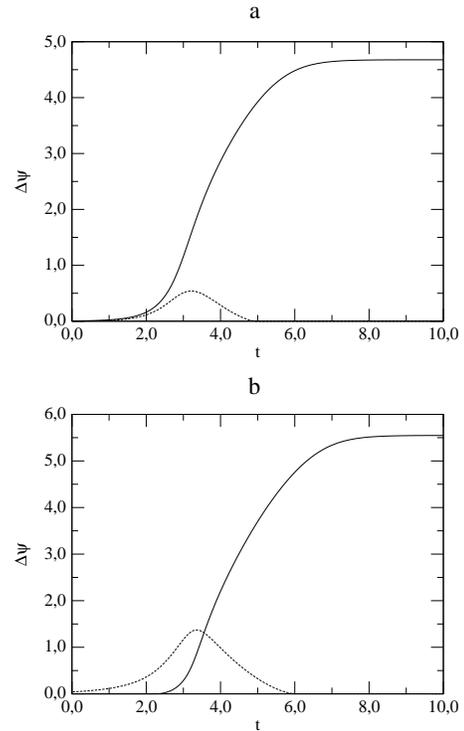,height=10cm,clip=t}}}
\caption{Individual growth rates $\Delta \psi_1(t)$ and
$\Delta \psi_2(t)$ of the two fingers for the modified minimal model
with $\lambda=\frac{1}{2}$ and $\epsilon=0.1$, for
two different initial conditions showing successful competition.
For the (a) case the finger which initially has larger
growth rate (and larger length too) wins the competition.
For the (b) case the finger which initially has lower
growth rate (and lower length too) wins the competition,
in opposition to the evolution with the regularized dynamics
(small surface tension).}
\end{figure}
of the initially
small finger is higher than the flux of the other finger and finally 
the flux of the initially large finger reaches zero:
it has been suppressed from the competition. 
This is opposite to what it would expected \it a priori \rm from simple 
physical considerations. In fact, 
one would expect that for well developed fingers the
larger one wins the competition, at least if
the distance between the two finger tips is large before  
the process begins. This anomalous competition dynamics is illustrated in
figure 13b, where it can be seen that initially only one finger has
a finite $\Delta \psi_1$, and it grows until a second finger 
develops and begins to grow, as indicated by the appearance of a
non-zero $\Delta \psi_2$. The second finger grows together with
its growth rate and surpasses the first one, 
which is eventually suppressed from the competition process. This
is indicated by $\Delta \psi_1$ going to zero. This example is important
as a case where there is successful competition (finger coalescence)
to the Saffman-Taylor asymptotic solution but with a completely 
wrong dynamics. In fact it can be seen that the zero surface tension 
evolution departs from the regularized trajectory much before the small
finger takes over the competition (through the impact of a daughter
singularity \cite{Paune02}). The winning finger with the regularized
dynamics is thus the losing finger with the zero surface tension one.

Again, in the limit $\epsilon \rightarrow 0$ these phenomena
appear even more dramatically, as a consequence of the structural
instability of the minimal model. In this limit, for a
$O(-\ln \epsilon)$ time we will observe two unequal fully
developed fingers advancing with a fixed tip distance, but
eventually the presence of finite $\epsilon$ will 'activate'
the competition process and one of
the two fingers will reduce its growth rate until fully
suppressed from the competition. If $\alpha''(0)>0$ the suppressed
finger will be the small one, but if $\alpha''(0)<0$ the
dynamically suppressed finger will be the large one.

\subsection{Comparison with the regularized dynamics.}

In order to compare the $d_0=0$ dynamics with the physical case 
of $d_0 \neq 0$, as we have done
for the $\epsilon = 0$ case, we use the construction introduced in 
Sect. \ref{comp1}. Consider a one dimensional set of initial 
conditions ($t=0$) of the form Eq. (\ref{fepsilon}) surrounding 
the planar interface (PI) fixed point $\alpha'=0$,
$\alpha''=0$, for a fixed $\lambda$ and $\epsilon$. We choose the 
points of this set 
infinitesimally close to $\alpha=0$, and therefore the interface is in the 
linear regime. The $d_0 \neq 0$ time evolution from 
$t=-\infty$ to $t=\infty$ of this set
spans a compact two dimensional phase space and defines a 
dynamical system $S^2(d_0)$,  
embedded in the infinite dimensional $S^\infty(d_0)$. 
We define the dynamical system  $S^2(0^+)$ as the limit of
$S^2(d_0)$ for $d_0 \rightarrow 0$, intersecting the one 
dimensional set at $t=0$.

Since in the linear regime the regularized problem for vanishingly small $d_0$
converges regularly to the $d_0=0$ solution, then the dynamical 
system $L^2(\lambda,\epsilon)$ of $d_0=0$ and $S^2(0^+)$ must
be tangent at the PI fixed point, $\alpha=0$.
According to solvability theory the selected width $\lambda$
of the ST finger in the limit $d_0 \rightarrow 0$ is $\lambda=1/2$. Then,
$L^2(1/2,\epsilon)$ and $S^2(0^+)$ must intersect at 1ST(R), 
$\alpha'=0,\alpha''=1$ and  1ST(L), $\alpha'=0, \alpha''=-1$.
In addition, from Ref. \cite{Siegel96b} it follows that the single-finger
solution with $d_0 \rightarrow 0$ converges uniformly to the 
the time-dependent Saffman-Taylor finger solution 
with $\lambda=1/2$.
This implies that $L^2(1/2,\epsilon)$ and $S^2(0^+)$ intersect 
not only in the fixed points that they have in common, but also 
on the line $\alpha'=0$ (that corresponds to the time-dependent ST solution).
For the dynamical system $S^2(0^+)$, 
the basins of attraction of 1ST(R) and 1ST(L) are 
two-dimensional and finite, and therefore there must be at least one 
separatrix trajectory between the two basins. This separatrix must end at a 
saddle fixed point (which does not exist
in the phase portrait of the  $d_0 = 0$ solution). It is reasonable 
to assume that this fixed point is the double ST finger fixed point (2ST).
Thus, the topology of the flow defined by the dynamical system
 with $d_0=0$, $L^2(1/2,\epsilon)$ is different from the 
flow of the dynamical system $d_0 \rightarrow 0$, $S^2(0^+)$:
the flow 
for the regularized problem contains a trajectory and a fixed point
that it is not contained in the flow defined by the modified minimal model, 
the trajectory starting
at the planar interface PI fixed point and ending up at the 2ST fixed point.
The phase flow of the modified minimal model with $d_0=0$ is qualitatively 
different from the phase flow of the regularized problem, 
$d_0 \rightarrow 0$, and therefore the solution Eq. (\ref{fepsilon})
is unphysical in a global sense, what is to say, when a sufficiently large
set of initial conditions (spanning evolutions towards 1ST(R) and 1ST(L)) 
is considered simultaneously.

It is useful to define a projection of the trajectories from the 
regularized two-dimensional space $S^2(0^+)$ onto the $d_0=0$ 
$L^2(1/2,\epsilon)$ space in order to make possible 
not only the comparison between the topology of the two flows but also 
the comparison between individual trajectories. 
It is obvious that there are many different
possible definitions, and the projection of the regularized
flow $S^2(0^+)$ onto $L^2(1/2,\epsilon)$ is in that sense not  
unique, but the following discussion should not depend on the particular
definition of the projection. One possible definition of a useful
projection is the following \cite{Paune01}: 
given a point $\zeta$ of $S^2(0^+)$ and a point $\alpha$ of 
$L^2(1/2,\epsilon)$, we define a 'distance' $D_\zeta(\alpha)$ 
between $\zeta$ and $\alpha$, that is, a distance between an interface within
$S^2(0^+)$ and an interface within $L^2(1/2,\epsilon)$.
One possible choice of $D_\zeta(\alpha)$ is 
the area enclosed or \it trapped \rm between the two interfaces. 
Then, we define the projection $\alpha_\zeta$ of $\zeta$ 
(that corresponds to an
interface solution of the problem with $d_0 \neq 0$) onto the $d_0=0$
phase space $L^2(1/2,\epsilon)$
 as the value of  $\alpha$ that minimizes
the 'distance' $D_\zeta(\alpha)$ between $\alpha$ and $\zeta$,
or equivalently, that minimizes the 'distance' between the two
interfaces,
with the restriction that the average position of the two interfaces 
is the same to ensure that the projection satisfies mass conservation.
Then, once we have defined a projection $\alpha_\zeta$ of a point 
of $S^2(0^+)$ it is straightforward to obtain the projection
of a particular trajectory $\zeta(t)$ in $S^2(0^+)$, applying the 
definition of $\alpha_\zeta$ to any point of the trajectory.
This projection of a trajectory will be denoted by 
$\alpha_{\zeta_0}(t)$.

The projection on the $d_0=0$ phase portrait of the $d_0 \rightarrow 0$
separatrix trajectory  as $\epsilon \rightarrow 0$ will approach 
the time-dependent 2ST trajectory for $\epsilon=0$, the line $\alpha''=0$, but a broad
set of initial conditions of $\epsilon \rightarrow 0$ located below 
the projection of the 
separatrix (that is, with $\alpha'' < 0$) will be attracted 
by the 1ST(R) fixed point under the 
$d_0=0$ dynamics instead of being attracted by the 1ST(L) fixed point
as their position with respect the separatrix implies \cite{Paune01}. Thus, the evolution 
under the $d_0=0$ dynamics of this set of initial conditions 
will be not only quantitatively but also qualitatively incorrect, in the 
sense that the finger that wins the competition is the one that 
in the regularized dynamics loses. An example of this was given in Fig.13b. 
More details will be presented elsewhere \cite{Paune02}.

The motivation for the introduction of a finite $i\epsilon$
term to the minimal model Eq. (\ref{eq:map}) was to obtain 
an unfolding of 
its non-hyperbolic fixed point structure, but in this section
we have shown that the introduction of $i\epsilon$
has failed to unfold the phase portrait to the expected physical topology,
the saddle-point structure of the regularized problem. 
It has dramatically changed the topology of the flow obtained
for $\epsilon = 0$, but the flow for $\epsilon \neq 0$
does not have the expected structuraly stable flow of the physical problem: an unstable
fixed point, two stable fixed points and one saddle fixed point.
 Instead of this, the evolution of Eq. (\ref{fepsilon}) with 
$\epsilon \neq 0$  
presents finite-time singularities for a non-zero measure set of initial conditions. 
This can be understood as 
a consequence of the absence of the 2ST saddle point, 
which controls the competition regime. Without
this fixed point the separatrix trajectory necessary to separate
the basins of attraction of ST(L) and ST(R) is not present and the only
possible way to split the flow is through the existence of finite time 
singularities. This is not a particularity of the mapping Eq. (\ref{fepsilon}) 
but a more general feature of $d_0=0$ solutions. 
Below we will prove that, within the N-logarithms class,
finite $\epsilon$ implies finite-time singularities in the evolution of 
a non-zero measure set of initial conditions
(see Sect. \ref{cusps}). Besides the existence of finite-time singularities
we have seen that, unlike the case $\epsilon=0$ solutions exhibiting successful
competition are possible with $\epsilon \neq 0$ for $\lambda=1/2$. However, 
part of those evolutions are unphysical in the sense the winning finger 
may differ from the one with the regularized dynamics.

\section{Generalization to higher dimensions}
\label{General}

This section is devoted to the study of solutions that define a
dynamical system of dimensionality greater than two. We will show
that the discussion of previous sections is not peculiar  
of dimension $2$ but is extendable to higher dimensions.

\subsection{Non-axisymmetric fingers}

The solutions that have been studied in the previous sections, 
Eq. (\ref{eq:map}) and Eq. (\ref{fepsilon}), have two pole-like singularities
$\omega_{1,2}$ located at $\omega_1=1/\alpha$ and 
$\omega_2=-1/\alpha^*$. The property $\omega_1=-\omega_2^*$
reduces the dimensionality of the dynamical system to two and also
forces the axisymmetry of the fingers. If the singularities
$\omega_{1,2}$ are not related, then  
the phase space has one additional dimension and the fingers 
are not axisymmetric. 

We will now study the ansatz 
\begin{eqnarray} 
\label{non-axi}
f(\omega,t)=-\ln\omega + d(t) + 
(1-\lambda)\ln(1-\alpha_1(t)\omega) \nonumber \\
(1-\lambda)\ln(1-\alpha_2(t)\omega) 
\end{eqnarray} 
where $\alpha_j(t)=\alpha'_j(t)+i\alpha''_j(t)$ are two complex
quantities satisfying $|\alpha_j|<1$. This ansatz is an exact solution of 
Eq. (\ref{eq:evol}) and it can be proved that the evolution is free
of finite time singularities.

From the evolution equations for $\alpha_{1,2}$ in the case $\lambda=1/2$
it is found that the dynamics satisfy
\begin{equation}
\label{constant}
\frac{\alpha'_1\alpha'_2-\alpha''_1\alpha''_2}
{\alpha'_1\alpha''_2+\alpha''_1\alpha'_2}={\rm Const}.
\end{equation}
Writing $\alpha$ in polar coordinates, $\alpha_j=r_j e^{i\theta_j}$,
this constraint reduces to the simpler form
\begin{equation}
\label{constant1}
\theta_1+\theta_2={\rm C}.
\end{equation}
The constant C depends on the initial condition, but its value can
be fixed arbitrarily using the property of rotational invariance of
the ansatz Eq. (\ref{non-axi}): the transformation $\alpha_1 \rightarrow
\alpha_1 e^{i\phi}$ and $\alpha_2 \rightarrow \alpha_2 e^{i\phi}$ is
equivalent to a translation of the interface a distance $\phi$ in the $y$
axis direction. For the minimal class studied previously
(Eq. (\ref{eq:map})) the value of C was ${\rm C}=\pi$. The existence of
the constraint Eq. (\ref{constant1}) reduces the dimensionality of the
problem from four to three, with variables $r_1$, $r_2$ and
$\theta_1-\theta_2$, implying that the dynamical system is actually 
three-dimensional.

In order to study the dynamical system defined by the substitution
of the ansatz Eq. (\ref{non-axi}) into the evolution equation 
(\ref{eq:evol}) we introduce the variables $z=r\exp{i\theta}=\alpha_1+
\alpha_2$ and $\rho=\alpha_1\alpha_2$. Moreover, we use the arbitrariness
of C to choose the simpler value ${\rm C}=0$. Then,  
the dynamical system in the variables $(r,\theta,\rho)$ reads
\begin{eqnarray}
\label{sist3d}
\dot{r} & = & 4r\frac{1-r^2+\rho(2+r^2)\cos(2\theta)-3\rho^2}
{4-r^2} \nonumber\\
\dot{\rho} & = & 4\rho\frac{2(1-\rho^2)+r^2(\rho\cos(2\theta)-1)}
{4-r^2} \\
\dot{\theta} & = & -2\rho\sin(2\theta). \nonumber
\end{eqnarray}
With these  variables the axisymmetric case studied in previous sections 
is contained as a the particular case 
$\theta=0$. From the last equation it is immediate to see 
that $\theta$ decreases monotonically with time (since, for ${\rm C}=0$,
$\rho > 0$), and tends to zero
for long times, $\theta \rightarrow 0$, converging to the axisymmetric
case $\theta=0$. Therefore, we can conclude that the axisymmetric case 
is attracting the dynamics of the non-axisymmetric one 
and, asymptotically, its dynamics reduces to
the axisymmetric one. 

Therefore, we have shown that the dynamics of the minimal model is not
a particularity of a zero-measure subset but the general behavior
of the non-axisymmetric neighborhood of the original axisymmetric class.
This neighborhood has a non-zero measure within the three dimensional
system. Similarly, it can be shown that the basin of attraction
of the Saffman-Taylor finger for the ansatz Eq. (\ref{non-axi}) 
is also relatively small.\\

It is worth noting that non-axisymmetric fingers can be obtained
also in 2d. A conformal mapping describing non-axisymmetric fingers is
\begin{eqnarray}
f(\omega,t)=-\ln \omega + d(t) + (1 - \lambda + p + i\epsilon)
\ln (1 - \alpha (t) \omega) \nonumber \\
 + (1 - \lambda - p - i\epsilon)\ln (1 + \alpha^* (t) \omega)
\end{eqnarray}
where $0<p<1-\lambda$. However, the lack of axisymmetry caused by the 
introduction of $p$ does not change qualitatively the phase portraits
obtained for $p=\epsilon=0$, the continuum of fixed points present for $\epsilon=0$
is not removed by the introduction of a finite $p$ and the finite-time
singularities that appear for $\epsilon \neq 0$ are also present when $p\neq 0$.
Therefore, the results obtained for 2d in the previous sections are not
modified at all if we relax the condition of finger axisymmetry.

\subsection{Perturbations which change finger widths}

The second type of modification of the ansatz (\ref{eq:map}) we have studied
is the following. Consider
\begin{eqnarray}
\label{seleccio}
f(\omega,t)=-\ln \omega + d(t) + (1 - \lambda) 
\ln (1 - \alpha_1 (t) \omega) \nonumber \\
+ (1 - \lambda) \ln (1 -  \alpha_2 (t) \omega)\\
+ 2(\lambda - \lambda_s) \ln (1 -  \delta (t) \omega) \nonumber
\end{eqnarray} 
with initial conditions $\alpha_1 (0) = -\alpha_2^* (0) = \alpha (0)$,
$0 < \lambda,\lambda_s <1$ and $|\delta (0)| \ll 1$. Note that for 
$\delta (0)=0$ this ansatz reduces to the minimal model (\ref{eq:map}).
From substitution of this ansatz into the evolution equation (\ref{eq:evol})
it is obtained that Eq. (\ref{seleccio}) is a solution with $\lambda$ and $\lambda_s$
constants. From the dynamical equations it can be proved that 
the asymptotic configuration of this ansatz consists of one or two fingers, 
with asymptotic filling fraction equal to $\lambda_s$. But if 
$|\delta (0)| \ll |\alpha (0)|$ then the interface will be initially 
almost identical to the one obtained within the class (\ref{eq:map}) with 
the same $ \alpha (0)$
and $\lambda$, and its evolution will remain close to the one obtained with the
minimal class for a time that will increase with decreasing $|\delta (0)|$.
Therefore, given a small enough $|\delta (0)|$, starting from the planar 
interface a configuration with one or two fingers (depending on the 
initial conditions) of total width $\lambda$ will develop. 
Later on, as $|\delta|$ grows and approaches 1,
the total width will change from $\lambda$ to $\lambda_s$ 
for long enough time. The ansatz (\ref{seleccio}) thus
describes an interface that changes the filling fraction of
the fingers from $\lambda$ to $\lambda_s$. The same phenomenon 
will appear with any other of the solutions described in this 
paper (and in general in pole-like 
solutions) if a term of the type 
$2(\lambda - \lambda_s) \ln (1 -  \delta (t) \omega)$
is added, and in particular it will appear in the single 
finger configurations obtained setting $\alpha (0)$ 
and $\delta (0)$ purely imaginary. Note that in this case
if  $\delta (0)=0$ the ansatz describes exactly the 
time-dependent Saffman-Taylor finger.
This changing-width phenomenon of $d_0=0$ solutions
has been known for long
\cite{Howison86}, but it has been recently 
claimed \cite{Mineev97} to be responsible
of the known width selection observed with finite surface tension
both experimentally and numerically. 
The idea was that,
although solutions of arbitrary $\lambda$ exist in the absence 
of surface tension, these are unstable under some perturbations
which trigger the evolution towards the $\lambda = 1/2$ solution.
Since the present paper is basically emphasizing the unphysical 
dynamics of the idealized ($d_0 = 0$) problem, in direct contradiction with 
Ref. \cite{Mineev97}, we feel compelled to 
briefly comment on this respect here.
The basic argument
of Ref.\cite{Mineev97} is as follows, in terms of the parameterization
of the interface used by the author:  a term of the form $i\mu\phi$ in
the conformal mapping is always unstable under the substitution
$i\mu\phi \rightarrow \mu \ln(e^{i\phi}-\epsilon)$. The introduction
of such perturbation then leads to the $\mu=0$ case, which corresponds
to $\lambda=1/2$. In Refs.\cite{Casademunt98,Almgren98} it was pointed
out that, with the same degree of generality, equivalent perturbations
exist which lead to any desired $\lambda$, and therefore the conclusion
that $\lambda=1/2$ is the only attractor is incorrect.
It is argued\cite{Mineev98b} that the latter class of perturbations is
different form the former since they increase the number of logarithmic
terms in the conformal mapping and therefore modify the dimension of the
subspace of solutions. This objection is somewhat misleading since 
such partitioning of classes of solutions in terms of the number of
logarithms 
is arbitrary and not intrinsic. This can 
be seen by choosing a different
 reference region to conformally map the physical fluid. Instead of mapping
it into the semi-infinite strip \cite{Mineev97}, the mapping into the interior
of the unit circle avoids the confusion on the dimension of the subspace 
of solutions. Thus, the perturbation proposed in Ref. \cite{Mineev97} is 
equivalent to choosing $\lambda_s=1/2$ in the ansatz (\ref{seleccio}), but
it is manifest in this formulation that there is nothing special with this
particular choice of $\lambda_s$. 
Perturbations leading to any finger width $\lambda_s$ occur with
the same genericity. Therefore the instability of the point $\delta = 0$
is \it not \rm related to the steady state selection phenomenon.

\subsection{Finite-time singularities within N-logarithms solutions}
\label{cusps}

In this section we will prove that any solution of the N-logarithm class
\cite{Dawson94} that does not have only real constant parameters presents
finite time singularities, that is, it contains a non-zero measure set
of initial conditions which develop singularities at finite time.

Consider a conformal mapping function $f(\omega,t)$
\begin{eqnarray}
\label{generic}
f(\omega,t) = -\ln \omega + d(t) + 
(\Lambda_1 + i \epsilon) \ln(1-\alpha_1(t)\omega) \\ 
\nonumber + (\Lambda_2 - i \epsilon)  \ln(1-\alpha_2(t)\omega)  
\end{eqnarray}
where $\Lambda_1 + \Lambda_2 = 2(1 - \lambda)$, $\epsilon > 0$ and 
$\alpha_{1,2}$ are
complex with $|\alpha_{1,2}| < 1$. 
The mapping $f(\omega,t)$ must satisfy $\partial_{\omega}f(\omega,t)\neq 0$
for $|\omega| \leq 1$. If any zero $\omega_0$ of $\partial_{\omega}f(\omega,t)$
hits the unit circle $|\omega| = 1$ then the interface develops a cusp.
For the ansatz
(\ref{generic}) $\partial_{\omega}f(\omega,t)$ reads
\begin{eqnarray}
\label{d_generic} 
\partial_{\omega}f= -\frac{1}{\omega} 
- \frac{(\Lambda_1 +  i \epsilon) \alpha_1}{1 - \alpha_1 \omega}
- \frac{(\Lambda_2 -  i \epsilon) \alpha_2}{1 - \alpha_2 \omega}.
\end{eqnarray} 
Thus, the position of the zero $\omega_0$ of $\partial_{\omega}f(\omega_0,t)$
is 
\begin{eqnarray}
\label{zero} 
& \omega_0  = \frac{-(\Lambda_1 +  i \epsilon - 1) \alpha_1 -
(\Lambda_2 -  i \epsilon - 1) \alpha_2}{2 \alpha_1 \alpha_2 (2 \lambda -1)}
   \nonumber\\
&\pm  \frac{\sqrt{((\Lambda_1 +  i \epsilon
- 1) \alpha_1 + (\Lambda_2 -  i \epsilon - 1) \alpha_2)^2 -
4 \alpha_1 \alpha_2 (2 \lambda -1)}}{2 \alpha_1 \alpha_2 (2 \lambda -1)} 
\end{eqnarray}
If, for some value of $\alpha_{1,2}$, $|\alpha_{1,2}| \leq 1$, the zero $\omega_0$
is inside the unit circle, then the ansatz (\ref{generic}) will present
finite time singularities for some sets of initial conditions. Therefore, 
if $|\omega_0| < 1$ the interface will develop a cusp. 
Setting $\alpha_{1,2} = \alpha{\rm e}^{i \theta_{1,2}}$ and $\theta_2 - \theta_1 =
-2 \delta$ with $\delta \ll 1$ the position of the zero (keeping up to
linear terms in $\delta$) is:
\begin{eqnarray}
\omega_0 = {\rm e}^{-i \theta_2} \frac{\lambda \pm (1 - \lambda)}
{\alpha (2 \lambda - 1)} \nonumber \\
+ \frac{i \delta {\rm e}^{-i \theta_2}}{\alpha (2 \lambda - 1)} 
\left[ \Lambda_2 -1 -i \epsilon 
\pm \frac{\lambda - 1 + \lambda (\Lambda_2  -i \epsilon)}{1 - \lambda}\right] + 
{\rm O}(\delta^2)
\end{eqnarray}
and the modulus of the minus solution (the one with smaller modulus) reads
\begin{equation}
\label{modul}
|\omega_0| = \frac{1}{\alpha} [1 - \frac{\epsilon \delta}{1 - \lambda} 
+ {\rm O}(\delta^2)].
\end{equation}
In consequence, for $\alpha$ close to 1 we obtain $|\omega_0| < 1$, 
one of the zeros is inside the unit
circle in a finite neighborhood of $\alpha_1 = \alpha_2 = e^{i \theta}$.
Thus, the mapping (\ref{generic})
presents finite time singularities for some initial conditions independently
of the value of $\epsilon$ and $\Lambda_{1,2}$, and the measure of this 
set is non-zero.

Now we consider a generic mapping with $N>2$ logarithmic terms of the form:
\begin{eqnarray}
\label{Nlog}
f(\omega,t) = -\ln \omega + d(t) + \sum_{j=1}^N \gamma_j 
\ln (1 - \alpha_j (t) \omega)
\end{eqnarray}
where  $\gamma_j= \Lambda_j + i \Gamma_j$ are constants of motion with the
restriction $\sum_{j=1}^N \gamma_j = 2(1-\lambda)$. 
If we choose $\alpha_j = \alpha_1$ for $1 \leq j \leq k$
and $\alpha_j = \alpha_2$ for $k+1 \leq j \leq N$ we recover the mapping
(\ref{generic}). Therefore, the N-logarithm solution (\ref{Nlog}) contains
initial conditions that develop a cusp with this subset of $\alpha_j$, 
but the dimension of this subset is lower than the dimension of the phase
space, implying that the measure of the set is zero compared to the whole
phase space. To prove that the subset that develops a cusp is finite 
we choose now the following values for $\alpha_j$:
$\alpha_j = \alpha_1 + \eta_j$ for $1 \leq j \leq k$ and 
$\alpha_j = \alpha_2 + \eta_j$ for $k+1 \leq j \leq N$, with $|\eta_j| \ll 1$,
where $|\omega_0|<1$ if $\eta_j=0$. The equation
$\partial_{\omega}f(\omega,t) = 0$ reads
\begin{eqnarray}
\label{eqnNlog}
\frac{1}{\omega} + \sum_{j=1}^k \frac{\gamma_j(\alpha_1 + \eta_j)}
{1 - (\alpha_1 + \eta_j)\omega} + 
\sum_{j=k+1}^N \frac{\gamma_j(\alpha_2 + \eta_j)}
{1 - (\alpha_2 + \eta_j)\omega} = 0.
\end{eqnarray}
This equation (\ref{eqnNlog}) reduces to Eq. (\ref{zero})
if all $\eta_j=0$ and it has $N$ zeros if $\eta_j \neq 0$.
Defining $g(\omega) = \partial_{\omega}f(\omega,t)$ for $\eta_j=0$ and
$G(\omega,\vec{\eta}) = \partial_{\omega}f(\omega,t)$ for $\eta_j \neq 0$
then $G(\omega,\vec{\eta}) = g(\omega) + \delta G(\omega,\vec{\eta})$
where $|\delta G(\omega,\vec{\eta})| < K |\vec{\eta}|$ for $|\omega| < R$,
with $K$ and $R$ constants, and $g(\omega_0) = 0$. 
One zero $\omega'_0$
of $G(\omega,\vec{\eta})$ can be written $\omega'_0=\omega_0 + \delta \omega$,
and assuming $|\delta \omega| < C |\vec{\eta}|$ with $C$ constant the 
substitution of  $\omega'_0$ in $G(\omega,\vec{\eta}) = 0$ yields
\begin{equation}
\label{ijoquese}
g(\omega_0) + \left.\frac{\partial g}{\partial \omega}\right|
_{\omega_0} \delta \omega + 
\delta G(\omega_0,\vec{\eta})=0.
\end{equation}
The position of the zero is then:
\begin{equation}
\label{elzero}
\omega'_0=\omega_0 - \frac{\delta G(\omega_0,\vec{\eta})}
{\left.\frac{\displaystyle \partial g}
{\displaystyle \partial \omega}\right|_{\omega_0}}
\end{equation}
where $\left.\frac{\partial g}{\partial \omega}\right|_{\omega_0} \neq 0$.
Therefore, the zero $\omega'_0$ of Eq. (\ref{eqnNlog}) is inside a ball
of radius $o(|\vec{\eta}|)$ centered in $\omega_0$. If $|\omega_0| < 1$,
then choosing $|\vec{\eta}|$ small enough the zero will satisfy
$|\omega'_0| < 1$: 
in a neighborhood of $(\alpha_1, \alpha_2)$ at least one zero of 
$\partial_{\omega}f(\omega,t)$ is inside the unit circle, and the dimension
of this neighborhood will be the same of the phase space. 
So we can conclude that any mapping of the 
form (\ref{Nlog}) presents finite time singularities for some sets of initial
conditions of non-zero measure, provided that at least one pair of $\gamma_j$
has a non zero imaginary part.

Thus, the requirement that a mapping function of the form (\ref{Nlog})
is free of finite time singularities for any initial condition 
$\alpha_j(0)$ is fulfilled if and only if ${\rm Im} [\gamma_j]=0$,
$j=1,...,N$. But this restriction implies \cite{Mineev99} that 
for a wide range of initial conditions the
asymptotic configuration is a N-finger interface with unequal fingers 
advancing at a constant speed, a situation fully analogous to the
one discussed in Sect. \ref{Minimal}. Then, if a mapping of the form
(\ref{Nlog}) with ${\rm Im} [\gamma_j]=0$ is chosen, then the dynamical
system $L^{2N}(\gamma_j)$ will have non-hyperbolic fixed-points 
(continua of fixed points) and will lack the saddle-point
structure of the regularized problem. In order to completely remove
the continua of fixed points it is necessary to set 
${\rm Im} [\gamma_j] \neq 0$ \cite{Mineev99}, but in this case 
we will encounter
finite-time singularities and the saddle-point structure will not
be present anyway.

To sum up, we have shown that the features of the minimal model and its
extensions that make them unphysical (in a global sense) are not 
specific of their low dimensionality. 
The features that make the solutions studied
in previous sections ineligible as a physical description of low
surface tension dynamics for a sufficiently large class of initial 
conditions, are also present within the much more 
general N-logarithm family of solutions, and the conclusions
drawn in previous sections can be generalized to that class.

\subsection{Rigid-wall boundary conditions}
\label{rigid}

It is worth stressing here that the use of 
periodic boundary conditions throughout this study,
as opposed to the physically more natural rigid-wall boundary conditions, 
is not essential to the basic discussion. In connection with the discussion 
of multifinger steady solutions, this point 
was raised in Ref.\cite{Vasconcelos01} 
and addressed in Ref.\cite{Magdaleno01}. Here we will just recall that
the choice of periodic boundary conditions is not only the simplest 
in terms of symmetry and dimensionality, but it is the relevant one if 
one is interested in general mechanisms of finger competition in finger
arrays. In this sense, the study of the two-finger configurations 
in this paper refers to an alternating mode of two-finger 
periodicity in an infinite array of fingers, in the spirit 
of Ref.\cite{Kessler86b}. For finite size-systems 
one can also argue that rigid-wall boundary conditions are included as
a particular case of periodic boundary conditions in an enlarged 
system. That is, a channel with width $W$ with rigid walls in 
mathematically equivalent to a channel of width $2W$ with periodic boundary
conditions where auxiliary channel of with $W$ is constructed as the 
mirror image of the physical one. The competition of two fingers in 
a channel with rigid walls at a distance $W$ 
is in practice equivalent to a four-finger 
problem with periodic boundary conditions in a double channel.

The only subtle point which we would like to stress is the apparent 
degeneracy of the single-finger attractor into a left ST finger and 
a right ST finger, as already pointed out in Sect. \ref{study_ep},
and the possible relevance of 
this fact in connection with the saddle-point structure of the 
phase space flow.
This degeneracy is inherited from the trivial continuous 
degeneracy associated to translation invariance in the transversal 
direction, when periodic boundary conditions are assumed. In fact an
arbitrary shift in the transversal direction yields a physically 
equivalent configuration. When an initial condition is fixed, such 
continuous degeneracy is broken into two discrete spatial positions 
which are separated a distance $W/2$. The whole dynamical system is 
then invariant under translations of $W/2$. This is the reason why 
we only plotted a half of the disk in the phase portraits 
of section \ref{Beyond}
Technically, the resulting dynamical system must be defined 'modulo-W/2',
that is, identifying any configuration with the resulting of a $W/2$ 
shift. In the phase space defined by the variables $(\alpha',\alpha'')$
 one should identify 
any point with the resulting of a $\pi$ rotation. In this way the two 
single-finger attractors do correspond to the same fixed point. 
With this identification, the ST finger is not degenerate and the flow 
becomes topologically equivalent to the corresponding one in a channel
with rigid-wall boundary conditions. The two-finger configurations 
have thus the same structure, regardless of the type of sidewall 
boundary conditions. The flow starts at the PI fixed point and 
ends up at the 1ST fixed point. Between them there is a saddle point 
corresponding to the 2ST fixed point. This separates the flow in two 
equivalent regions, namely 'from the left' and 'from the right' of the 
saddle point. With zero-surface tension, the case of rigid walls 
exhibits the same problems, namely the occurrence of a (nontrivial) 
continuum degeneracy of multifinger solutions, and the existence of 
finite-time singularities.
The important point we want to stress is thus that all the general
conclusions drawn in this paper are valid if rigid-wall boundary 
conditions are considered.

\section{Dynamical Solvability. General discussion}
\label{discussion}

\subsection{The physics of zero-surface tension}
\label{physics}

The role of the zero surface tension solutions in the description of
the dynamics of the nonzero but vanishingly small surface tension problem 
is now clearer.
The $d_0=0$ dynamics is in general incorrect in a global sense, 
even if we choose solutions with the asymptotic width $\lambda$ given 
by selection theory.
However, they have an important place in the description of the
physical problem. It has been proved in 
Refs. \cite{Tanveer93,Siegel96a,Siegel96b} 
that the solutions with $d_0=0$ converge to the 
$d_0 \rightarrow 0$  during a time $O(1)$, before the impact  with the 
unit circle of the so-called \it daughter singularity \rm at time $t_d$. 
In practice 
 this implies that the $d_0=0$ dynamics
is not only correct in the linear regime (where $d_0$ acts as regular perturbation)
 but also quite deep into the nonlinear
regime. After $t_d$ nothing can be said \it a priori\rm:  
as we have shown in the present paper, there are regions of the $d_0=0$
phase space corresponding to smooth interfaces that are physically wrong, 
but other regions are a good
description of the evolution with finite (but very small) surface tension. 
For instance, in the neighborhood of the time-dependent Saffman-Taylor finger
(the line $\alpha'=0$ in the solutions (\ref{eq:map}), (\ref{fepsilon}))
the $d_0=0$ evolution is qualitatively correct for finite surface tension,
and even quantitatively correct in the limit $d_0 \rightarrow 0$
(for $\lambda=1/2$).
However, a question remains open: given a $d_0=0$ evolution smooth for all time
and consistent with the results of selection theory, 
is it the limit of a 
$d_0 \rightarrow 0$ evolution? This question can be explored numerically
and is the subject of a forthcoming paper \cite{Paune01}. Generally speaking
the conclusion is that exact solutions including evolution of 
two different fingers which are compatible with MS theory, that is, 
evolving to a single finger with the width predicted by selection theory, 
and which do not exhibit any kind of singularity in the interface shape, may 
be dramatically affected by surface tension. The outcome of the competition,
that is, which one of the two competing fingers will survive at the end, 
when an infinitesimally small surface tension in introduced may be the 
opposite of that of the zero surface tension case. This may happen in 
situations where fingers are significantly different from each other and
is not an instability of a particular trajectory, but a generic behavior
in a finite (non-zero measure) range of initial conditions within the 
integrable class. For that region of phase space, it is clear that the 
dynamics of finger competition is completely wrong for the class of integrable 
solutions. Nevertheless, there is also a class of initial conditions which
have a qualitatively correct evolution including 'successful' finger 
competition in the sense defined in sections above (this possibility was 
incorrectly excluded in Ref.\cite{Magdaleno98}, where the analysis was 
based on $\epsilon=0$). Although strict convergence of the regularized 
solution to 
the idealized one may not occur in these cases, 
the quantitative differences may be moderately
small. Actual convergence of some type can only be expected at most 
when there is only one finger along the complete time evolution.
In summary, according to this scenario there are basically four classes of
initial conditions within the most general integrable solutions, once 
those \it a priori \rm incompatible with selection theory are excluded, namely
(i) finite-time singularities forward or backward (or both) in time;
(ii) asymptotically correct ST finger with wrong dynamics (the incorrect 
finger wins);
(iii) asymptotically correct ST finger with qualitatively correct evolution
(the correct finger wins although shapes may differ during a transient); and
(iv) (unphysical) evolution towards multifinger fixed points. It has to be added that, 
all of the above solutions plus those which are incompatible with selection
theory are qualitatively and quantitatively correct in the limit of small
surface tension, until a time of order
one which is always in the deeply nonlinear regime.

As a general consideration it is worth remarking that fingers 
emerging from the instability of the planar interface when this 
is subject to noise are necessarily in the range of dimensionless 
surface tension of order one. A simple way to argue this point is that
it is precisely surface tension which selects the size of the emerging 
fingers, since the fastest growing mode is that in which both stabilizing 
and destabilizing forces are of the same order. In these cases, surface 
tension is felt necessarily in the linear regime, and the usefulness of 
the zero surface 
tension solutions in the early stages of the evolution 
is obviously more limited.

Finally, from a physical point of view it is appropriate to recall that
the presence of noise does modify the general picture of the fingering 
dynamics in the limit of small surface tension, as pointed out in 
Ref.\cite{Kessler01}. Although the ST finger is the universal attractor
of the problem, the \it linear \rm basin of attraction decreases 
with dimensionless surface tension. In practice this implies that 
the interface approaches the ST finger but when it gets too close, noise
triggers its nonlinear instability and the interface makes a long excursion
(typically a tip splitting) before approaching again the ST finger. 
The considerations made in this paper concerning the limit of small 
surface tension thus imply that noise must be taken as sufficiently small.
A careful discussion of the effects of noise, particularly in numerical 
simulation of very small surface tension will be presented 
elsewhere \cite{Paune01}.

\subsection{A Dynamical Solvability Scenario}
\label{Dynsol}

In Ref.\cite{Magdaleno98} we pointed out for the first time the dynamical 
implications of the MS analysis when extended to multifinger 
fixed points. The idea of the Dynamical Solvability Scenario (DSS) was already
latent in that discussion.  
We pursued this extension of the steady state
selection problem explicitly in Ref.\cite{Casademunt00,Magdaleno99}, 
where we found
that, in direct analogy to the single-finger case, the introduction of 
surface tension did select a discrete set of multifinger stationary states,
in general with coexisting unequal fingers. Here we would like to 
discuss in what sense that analysis does provide a Dynamic Solvability 
Scenario.

Before doing that, let us briefly consider an alternative view of a
possible DSS proposed by Sarkissian and Levine \cite{Sarkissian98}.
In Ref.\cite{Sarkissian98}, it was explicitly discussed with examples 
that exact solutions 
of the zero-surface tension problem did behave differently from numerical 
integration of the small surface tension problem. At the end, the authors 
speculated with the possibility that surface tension could play a 
selective role in the sense that it could basically pick up the physically
correct evolutions out of the complete set of solutions without surface 
tension, in direct analogy with the introduction of a small surface tension
selecting a unique finger width out of the continuum of stationary solutions.
Since the class of nonsingular integrable solutions is indeed vast and 
infinite-dimensional, it is not unreasonable to expect that one could 
approximate any particular evolution with finite surface tension with 
one of those solutions for all time. However, as recently pointed out in 
Ref.\cite{Kessler01}, there is no simple way to determine which of those
solutions is selected by any macroscopic construction. 
Furthermore, even if this were possible, one should still face the rather
uncomfortable fact that the base of solutions defined by the superposition 
of logarithmic terms in the mapping, would itself correspond to unphysical 
(nonselected) solutions, as we have seen throughout this paper. 
Indeed, an initial
condition defined exactly by a finite number of logarithms would have
to be replaced in general by a solution with an infinite number of 
logarithms as the 'selected' solution which the (small) 
finite surface tension system
tracks.

From a more general point of view, a dynamical selection principle 
understood as 'selection of trajectories' has an important shortcoming 
when considered within the perspective of a broader class of interfacial
pattern forming systems. In fact, the solvability theory of steady state
selection has turned into a general principle because its applicability to
a large variety of systems, most remarkably in the context of dendritic 
solidification\cite{Langer87,Pelce88,Kessler88,Brener91,Gollub99}. However, it 
is only for Laplacian growth problems that exact time-dependent solutions 
are known explicitly, so there would be no hope to extend the above 
DSS as a general principle to those other problems. 

The DSS we propose here has a weaker form but it is susceptible of 
generalization to other interfacial pattern forming systems.
The basic idea can be best expressed in similar words to those recently 
used by Gollub and 
Langer \cite{Gollub99} to describe solvability theory in a general 
context.
They have nicely synthesized the singular role 
of surface tension in the language of dynamical systems as to 'whether 
or not there exists a stable fixed point'\cite{Gollub99}.  
In this context, our DSS extends the (static) solvability scenario 
in the sense 
that the singular role of surface tension is precisely to guarantee 
the existence of multifinger fixed points with a saddle-point 
(hyperbolic) structure. We have seen that the continuum of multifinger 
fixed points is directly related to a nonhyperbolic structure of the 
equal-finger fixed points. They imply directions in phase space were
the flow is marginal, and this is so to all derivative orders.
While in the traditional solvability scenario the
introduction of surface tension does isolate a stable fixed point
(a continuum of single-finger fixed points turns into a stable one and
a discrete set of unstable ones), 
now it isolates multifinger saddle points out of continua of multifinger
solutions, as discussed in Ref.\cite{Magdaleno99,Casademunt00} 
(a continuum of n-finger fixed points turns into a hyperbolic fixed point 
with stable and unstable directions, and a discrete set of unstable ones).
Since the saddle fixed 
points are defined by the degenerate n-equal-finger solutions, the 
stable directions of the saddle-point are directly related to the 
stable directions of the single-finger fixed point, while the unstable 
directions correspond to all perturbations which break the n-periodicity
of the equal-finger solution. 
The most important stable and unstable directions,
however, are those depicted in the two-dimensional phase portraits discussed
in the above section, namely the 'growth' direction connecting the planar 
interface and the n-finger fixed point, and the 'competition' direction 
connecting the n-finger fixed point to the single-finger fixed point
\cite{footnote2}.
Notice that arrays of fingers emerging from the morphological instability
of the planar interface are relatively close of n-periodic solutions as 
long as the noise in the initial conditions is weak and white, which 
guarantees that the most unstable (fastest growing) mode dominates in 
the early nonlinear regime. In these conditions, the system feels the 
attraction  
to the corresponding n-equal finger fixed point. This stage is what we 
called 'growth'. When the fingers are relatively large they start to 
feel the deviations form exact periodicity and start the 'competition' 
process.

Note that, despite the formal analogy to the single-finger solvability 
theory, 
the reference to a the restoring of multifinger hyperbolicity by surface 
tension as \it dynamical \rm solvability scenario is fully justified.
Indeed, the local structure of the multifinger fixed point has a dramatic 
impact on the global (topological) structure of the phase space flow, 
as we have seen in simple examples. The existence of a small but finite 
surface tension thus determines a global flow structure and it is in this 
sense that it 'selects' the dynamics of the system. 

The possibility of extension of this analysis to other interfacial 
pattern forming problems relies on the existence of a continuum of 
unequal  multifinger 
stationary solutions with zero surface tension. The fact that in the 
ST case the existence of those can be associated to a simple relationship
between screening due to relative tip position and relative finger width
(that is, a slower areal growth rate of the screened finger is compensated
by its smaller width, resulting in an equal tip velocity), one is tempted
to conjecture that similar classes of solutions must exist in other 
problems, for instance in the growth of needle crystals in the 
channel geometry \cite{Brener91}. Although this point should be more
carefully addressed, it seems reasonable to expect that a DSS 
as presented above could be generalizable, to some extent, to other physical systems.

\section{Summary and conclusions}
\label{Concl}

We have developed a Dynamical Systems approach to study the dynamics
of the Saffman-Taylor problem, basing the analysis on the 
zero surface tension solutions. A minimal model has been analyzed, 
and from its phase flow we have concluded that 
it is unphysical. A detailed study of a perturbation of the
minimal model within two dimensions has yielded the same conclusion.
The unphysical behavior of zero surface tension solutions is a
consequence of the non-hyperbolicity of the multifinger fixed points 
of the finite-dimensional dynamical system that 
they define, opposed to the saddle point structure of the regularized
problem. Perturbations of the minimal model to higher dimensions
confirm the generality of the conclusions reached in two-dimensional models. 
We have proved that the N-logarithms class of solutions presents
finite-time singularities if the continua of fixed points are
totally absent. From the analysis of zero surface tension solutions 
we conclude that they are unphysical in a global sense, when sufficiently
large classes of initial conditions are considered simultaneously, 
because they lack the correct topology of the physical flow,  
structured in terms of a saddle-point connection between the unstable 
and the stable fixed points.
This does not exclude that,
for some sets of initial conditions, the zero surface tension dynamics 
might be correct, not only qualitatively but even quantitatively,
but it is not possible in practice to know it \it a priori \rm by
any simple means.
We have illustrated with several examples that although the asymptotic
behavior may be correct (evolution towards a single ST finger) the 
intermediate dynamics may be completely wrong, or even physically 
meaningless, such as for the existence of interface crossings. 
We have also illustrated the sensitivity to initial conditions when 
approximating physically relevant situations with different integrable 
solutions.
As a by-product we have also obtained some practical results concerning 
zero-surface tension dynamics which may be relevant to Laplacian growth 
problems, for instance in relation to the interplay of screening 
effects and finger widths. We have introduced precise definitions of 
'growth' and 'competition'. 
With the proper definition 'successful' competition, 
we can state for instance that, in the absence of surface tension, 
narrow fingers do compete more efficiently than wide ones. 
We have also found explicit solutions which lead to finite-time 
interface pinch-off in the stable configuration of the problem.

The detailed comparison of the dynamics with zero and non-zero but very small
surface tension requires a careful numerical study and can be analyzed 
in terms of the daughter singularities formalism developed in Refs.
\cite{Siegel96a,Siegel96b}. As a matter of fact it can be shown that the
zero surface tension problem and the vanishingly small surface tension 
regularization differ dramatically even in regions where the former is 
nonsingular, in the sense that non-zero measure regions of phase space have 
a different outcome of the competition (namely, which one of two 
competing fingers survives) in the two cases. A detailed study of this 
point will be presented elsewhere \cite{Paune01}.

Finally, we propose a Dynamical Solvability Scenario relevant in principle 
not only for viscous fingering problems but also of applicability to other
pattern forming  problems. Within this DSS the
role of surface tension as a singular perturbation is to
isolate multifinger saddle points out of the continua of
multifinger fixed points, as shown previously in 
Ref.\cite{Magdaleno99,Casademunt00}. 
This extends the traditional solvability theory applied
to steady state selection, where surface tension did 
also isolate a unique (stable) hyperbolic 
fixed point out of a continuum of nonhyperbolic ones. In that case the 
isolated fixed point was the global attractor of the problem. In the 
present extension, the introduction of surface tension does isolate 
a unique n-equal
finger fixed point out of each continuum of n-finger fixed points, with 
both stable and unstable directions. By restoring this saddle point 
local structure the topology of the phase space flow is
modified, so the introduction of surface tension has a deep impact on the global 
phase-space structure of the dynamics.
It is in this sense that this scenario can be considered as a \it 
dynamical \rm solvability theory.

\section*{Acknowledgments}

We acknowledge financial support from the Direcci\'on General de 
Ense\~nanza Superior (Spain), Project No. BXX2000-0638-C02-02, and from the
EU TMR network project ERB FM RXCT 96-0085. E. Paun\'e also acknowledges
financial support from the Departament d'Universitats, Recerca i Societat
de la Informaci\'o (Generalitat de Catalunya).

\end{multicols}


\begin{references}
\bibitem{footnote0}
In principle, the compactness of the resulting dynamical system could be
troublesome for the present analysis due to
the unboundedness of the original problem. Unlike what happens for instance 
for low viscosity contrast Hele-Shaw flows, 
where pesistent dynamics of all regions of the 
interface may occur \cite{Paune02}, the dynamics of high 
viscosity contrast is sufficiently 
simple, that is, compatible with low-dimensional dynamical systems in a 
compact space, that no such subtelties are relevant to the cases studied
in this paper.
\bibitem{footnote1}
For simplicity of the discussion we abusively use the same notation 
to refer to the dynamical system and to the manifold
on which it is defined.
\bibitem{modes}
The amplitudes of the first two modes $k=1,2$ are $\delta_1=2(1 - \lambda)
\alpha'' + \epsilon \alpha'$ and $\delta_2=2\epsilon \alpha'\alpha''
+ (1-\lambda)(\alpha''^2-\alpha'^2)$.
\bibitem{footnote2}
The observed hierarchical elimination of fingers in stages which reduce them
in number approximately by a factor one half \cite{Kessler86b,Curtis89}
 could have here direct implications on the structure of the connections 
among the n-equal-finger fixed points.
\end{references}
\end{document}